\documentclass{JHEP3} 

\JHEPspecialurl{http://jhep.sissa.it/JOURNAL/JHEP3.tar.gz}

\usepackage{epsfig,multicol,bbm}
\usepackage{amssymb,latexsym}
\usepackage{amsmath}
\usepackage{multirow}
\usepackage{amsfonts}
\usepackage{bm,euscript,array,cite}

\newcommand\fverb{\setbox\fverbbox=\hbox\bgroup\verb}
\newcommand\fverbdo{\egroup\medskip\noindent%
			\fbox{\unhbox\fverbbox}\ }
\newcommand\fverbit{\egroup\item[\fbox{\unhbox\fverbbox}]}
\newbox\fverbbox

\title{Dimensional Reduction of the Heterotic String over nearly-K\"{a}hler manifolds}

\author{Athanasios Chatzistavrakidis$^{1,2}$ and George Zoupanos$^1$
\\ $^1$Institute of Nuclear Physics, NCSR Demokritos, GR-15310 Athens, Greece\\
    $^2$Physics Department, National Technical University of Athens, GR-15780 Zografou Campus, Athens, Greece\\
	
\

E-mails: \email{cthan@mail.ntua.gr}, \email{zoupanos@cern.ch}}

\abstract{Our aim is to derive the effective action in four dimensions resulting by reducing dimensionally the ten-dimensional ${\cal N}=1$ heterotic supergravity coupled to ${\cal N}=1$ super Yang-Mills over manifolds admitting a nearly-K\"{a}hler structure. Given the fact that all homogeneous six-dimensional nearly-K\"{a}hler manifolds are included in the class of the corresponding non-symmetric coset spaces plus a group manifold, our procedure amounts in applying the Coset Space Dimensional Reduction scheme using these coset spaces as internal manifolds. In our examination firstly the rules of the reduction of the theory over a general six-dimensional non-symmetric manifold are stated and subsequently a detailed case by case analysis is performed for all the three non-symmetric coset spaces. For each case the four-dimensional scalar potential is derived and the corresponding nearly-K\"{a}hler limit is obtained. Finally, we determine the corresponding supergravity description of the four-dimensional theory employing the heterotic Gukov-Vafa-Witten formula and results of the special K\"{a}hler geometry.}

\begin{document}

\newcommand{\nn}{\nonumber}
\newcommand{\miso}{\frac{1}{2}}
\def\beq{\begin{equation}}
\def\eeq{\end{equation}}
\def\bea{\begin{eqnarray}}
\def\eea{\end{eqnarray}}
\def\mc{\mathcal}
\newcommand{\m}{\mathbf}
\newcommand{\fet}{\frac{1}{3}}
\newcommand{\fdt}{\frac{2}{3}}
\newcommand{\ftt}{\frac{4}{3}}
\def\w{\wedge}
\def\olra{\overleftrightarrow}
\def\vf{\varphi}
\def\g{\gamma}
\def\1{{G_2}/{SU(3)}}
\def\2{{Sp_4}/{SU(2)\times U(1)}}
\def\3{{SU(3)}/{U(1)\times U(1)}}
\def\bi{\begin{itemize}}
\def\ei{\end{itemize}}
\def\t{\tilde}

\newpage

\section{Introduction}

The heterotic string \cite{Gross:1984dd} has always been considered one of the most promising versions of the string theory in the prospect to find contact with low-energy physics studied in accelerators, mainly due to the presence of the ten-dimensional ${\cal N}=1$ gauge sector. Upon compactification the initial $E_8\times E_8$ gauge group can break to phenomenologically interesting Grand Unified Theories (GUTs), where the standard model could in principle be accommodated\footnote{The case of the $SO(32)$ gauge group has limited phenomenological viability therefore we shall hereby focus on the $E_8\times E_8$ gauge group.}. Moreover, the presence of chiral fermions in the higher-dimensional theory serves as an advantage in view of the possibility to obtain chiral fermions also in the four-dimensional theory. Finally, the original supersymmetry provided the hope that using appropriate manifolds to describe the extra dimensions could survive, not enhanced, in four dimensions. In order to find contact with the minimal supersymmetric standard model, the non-trivial part of this scenario was to invent mechanisms of supersymmetry breaking within the string framework.

The task of providing a suitable compactification which would lead to a realistic four-dimensional theory has been pursued in many diverse ways for more than twenty years. The realization that Calabi-Yau (CY) threefolds serve as suitable compact internal spaces in order to maintain an ${\cal N}=1$ supersymmetry after dimensional reduction from ten dimensions to four \cite{Candelas:1985en} has led to pioneering studies in the dimensional reduction of superstring models \cite{Witten:1985xb},\cite{Derendinger:1985kk}. However, in CY compactifications the resulting low-energy field
theory in four dimensions contains a number of massless chiral
fields, known as moduli, which correspond to flat directions of the effective
potential and therefore their values are left undetermined.

The attempts to resolve the moduli stabilization problem have led to the study of compactifications with fluxes (for a review see e.g. \cite{Grana:2005jc}). In the context of flux compactifications the recent developments have suggested the use of a wider class of internal spaces, called manifolds with $SU(3)$-structure, that contains CYs. Admittance of an $SU(3)$-structure is a milder condition as compared to $SU(3)$-holonomy, which is the case for CY manifolds, in the sense that a nowhere-vanishing, globally-defined spinor can be defined such that it is covariantly constant with respect to a connection with torsion and not with respect to the Levi-Civita connection as in the CY case. Manifolds with $SU(3)$-structure have been exploited in supersymmetric type II compactifications \cite{Behrndt:2004km} -\cite{Cassani:2009ck} as well as in heterotic compactifications \cite{LopesCardoso:2002hd} -\cite{Benmachiche:2008ma}.

An interesting class of manifolds admitting an $SU(3)$-structure is that of nearly-K\"{a}hler manifolds. The homogeneous nearly-K\"{a}hler manifolds in six dimensions have been classified in \cite{Butruille:2006} and they are the three non-symmetric six-dimensional coset spaces and the group manifold $SU(2)\times SU(2)$. In the studies of heterotic
compactifications the use of non-symmetric coset
spaces was introduced in \cite{Govindarajan:1986kb} -\cite{Castellani:1986rg} and recently
developed further in \cite{LopesCardoso:2002hd},\cite{Manousselis:2005xa}.
Particularly, in \cite{Manousselis:2005xa} it was shown that
supersymmetric compactifications of the heterotic string theory of
the form $AdS_4\times S/R$ exist when background fluxes and general
condensates are present. Moreover, the effective theories resulting from dimensional reduction of the heterotic string over nearly-K\"{a}hler manifolds were studied at zeroth order in $\alpha'$ in \cite{Chatzistavrakidis:2008ii}.

The quest of finding supersymmetric Minkowski vacua of the heterotic string theory with stabilized moduli appears to be generically a difficult problem. The vacuum structure of heterotic string vacua with internal manifolds possessing an $SU(3)$-structure has been studied in \cite{deCarlos:2005kh} -\cite{deAlwis:2008vk}. The general outcome of these approaches is that no satisfactory supersymmetric vacua exist where the moduli are stabilized. A different approach was adopted in \cite{Derendinger:2005ed}, where the authors aim to find stationary points of the potential where supersymmetry breaks in Minkowski space.

Last but not least it is worth noting that the dimensional reduction of ten-dimensional ${\cal N}=1$ supersymmetric gauge theories over non-symmetric coset spaces led in four dimensions to softly broken ${\cal N}=1$ theories \cite{Manousselis:2000aj}.

In this article we discuss the dimensional reduction of the
heterotic string in the case where the
internal manifold is a non-symmetric coset space admitting a nearly-K\"{a}hler structure. In section 2 we provide a brief reminder of the heterotic supergravity coupled to super Yang-Mills and discuss the basics of
manifolds with $SU(3)$-structure. In addition, we discuss homogeneous nearly-K\"{a}hler manifolds, thus specifying the internal spaces we are going to use, and we briefly present the Coset Space Dimensional Reduction scheme, which we shall employ in order to perform the dimensional reduction. In section 3 we present the
general reduction procedure that we follow and determine the
resulting four-dimensional Lagrangian. We also analyze in detail the four-dimensional potential arising from the gravity sector. In section 4 we apply the
previously found results in the case of all the six-dimensional non-symmetric coset spaces. We determine the corresponding four-dimensional scalar potential for each example and discuss how the nearly-K\"{a}hler limit can be obtained. Then it is argued that some terms of this potential can be identified with the soft supersymmetry breaking sector of a Minkowskian four-dimensional theory. In section 5 a supergravity description of the above examples from the four-dimensional point of view is attempted. We determine for each case the K\"{a}hler potential with the aid of the results of the special K\"{a}hler geometry and the superpotential by the heterotic Gukov-Vafa-Witten formula. Our conclusions appear in section 6. In the appendix A we present the basics of the coset space geometry which are used in the calculations of sections 3 and 4. Then in the appendix B we collect the necessary geometric data of the homogeneous nearly-K\"{a}hler manifolds for our purposes and we present the relevant tables of field decompositions for the dimensional reduction.

\section{General Framework}

In this section we provide our general framework by briefly
reviewing the field content and the Lagrangian of the ${\cal N}=1$ heterotic
supergravity coupled to ${\cal N}=1$ super Yang-Mills to fix our notation and
conventions. We briefly describe the basics of
the theory of manifolds with $SU(3)$-structure and we focus on the homogeneous nearly-K\"{a}hler manifolds, which we shall use for the dimensional reduction. Finally, we specify the expansion forms and we also give an account on the coset space dimensional
reduction, stating the basic ideas and results which will be used in
the following sections.

\subsection{The spectrum and Lagrangian in ten dimensions}

The fields of the heterotic supergravity coupled to super Yang-Mills, which is the low-energy
limit of the heterotic superstring theory, consist of the ${\cal N} =1, D=10$ supergravity multiplet which
contains the fields $g_{MN}, \psi_{M}, B_{MN}, \lambda, \varphi$,
(i.e. the graviton, the gravitino which is a Rarita-Schwinger
field, the two-form potential, the dilatino which is a Majorana-Weyl
spinor, and the dilaton which is a scalar), coupled to an ${\cal N}
=1, D=10$ vector supermultiplet which contains the gauge field
$A_{M}$ and the corresponding gaugino $\chi$. The field $B_{MN}$ is
an abelian two-form essential for the cancelation of anomalies in
string theory \cite{Gross:1985fr}. The only possible anomaly-free
gauge groups that can be coupled to ${\cal N}=1$ supergravity in ten
dimensions are $SO(32)$ and $E_8\times E_8$ \cite{Green:1984sg}, \cite{AlvarezGaume:1983ig}. In
the following we shall mainly focus on the second possibility, which
is more plausible for model building since it can lead to
phenomenologically interesting GUTs. However, the general discussion
holds for both gauge groups.

The corresponding ten-dimensional Lagrangian, in the Einstein frame, can be
written as $
\mc{L}=\mc{L}_b+\mc{L}_f+\mc{L}_{int},
$ \cite{Bergshoeff:1981um}, where the different
sectors of the theory are{\footnote{Here we use differential form
notation for the kinetic terms of the bosons, which will prove to be
useful in the course of the reduction.}} \bea
\hat{e}^{-1}\mc{L}_{b}&=&
-\frac{1}{2\hat{\kappa}^2}\biggl(\hat{R}\hat{*}\mathbf{1}+ \frac{1}{2}
e^{- \hat{\phi}}\hat{H}_{(3)}\wedge \hat{*}\hat{H}_{(3)}+\frac{1}{2}
d\hat{\phi}\wedge
                \hat{*} d\hat{\phi}+\frac{\alpha'}{2}e^{-\frac{\hat{\phi}}{2}}Tr(\hat{F}_{(2)}\w\hat{\ast}\hat{F}_{(2)})\biggl),\nn\\
 \hat{e}^{-1}\mc{L}_{f}&=&-\miso{\hat{\bar\psi}}_M \hat{\Gamma}^{MNP}
            D_N\hat{\psi}_P
                 - \miso{\hat{\bar\lambda}}\hat{\Gamma}^M D_M \hat{\lambda}-\miso
                    Tr({\hat{\bar\chi}}\hat{\Gamma}^M D_M \hat{\chi}),\nn\\
 \hat{e}^{-1}\mc{L}_{int}&=& e^{-\hat{\phi}/2}\hat{H}_{PQR}\biggl(\hat{\bar{\psi}}_M\hat{\Gamma}^{MPQRN}\hat{\psi}_N
                                    +6{\hat{\bar\psi}^P}\hat{\Gamma}^Q\hat{\psi}^R- \sqrt{2}\bar{\psi}_M\hat{\Gamma}^{PQR}\hat{\Gamma}^M \hat{\lambda}+
                                Tr({\hat{\bar\chi}}\hat{\Gamma}^{PQR}\hat{\chi})\biggl)\nn
                                            \\
                                 &-&\miso{\hat{\bar\psi}}_M \hat{\Gamma}^N \hat{\Gamma}^M
                            \hat{\lambda}
                                \partial_N\hat{\phi}+ e^{-\hat{\phi}/4}Tr\biggl(\hat{F}_{MN}( {\hat{\bar\chi}}\hat{\Gamma}^P \hat{\Gamma}^{MN}\hat{\psi}_P
                                    +{\hat{\bar\chi}}\hat{\Gamma}^P
                                    \hat{\Gamma}^{MN}\hat{\Gamma}_P\hat{\lambda})\biggl),
                                               \eea up to four-fermion terms.
We have placed hats in all the ten-dimensional fields to
distinguish them from their four-dimensional counterparts which will
appear after the reduction. The gamma matrices are the generators of
the ten dimensional Clifford algebra, hence we place hats on them
too, while those with more than one index denote antisymmetric
products of $\Gamma$s. $\hat{\kappa}$ is the
gravitational coupling constant in ten dimensions with
dimensions [length]$^4$; $\hat{e}$ is the
determinant of the metric, while $\hat\ast$ is the Hodge star
operator in ten dimensions. Finally $\alpha'$ is the Regge slope parameter and it has dimensions [length]$^2$.

The bosonic sector of the Lagrangian, ${\cal L}_b$, clearly involves the
Einstein-Hilbert action in ten dimensions, the kinetic term for the
higher-dimensional dilaton, the kinetic term for the gauge fields
and the corresponding one for the three-form. The three-form
$\hat{H}$ is sourced by the $B$-field plus additional corrections from
Chern-Simons forms related to the cancelation of anomalies. A more
detailed account on this point will be given in section 3.3. Let us also note that the Lorentz Chern-Simons form, which is added in order to cancel the gravitational anomalies, breaks supersymmetry and hence an introduction of a Gauss-Bonnet term in the Lagrangian is needed in order to restore it. However, we shall not discuss this term since it is not needed in the minimal supergravity Lagrangian.

In the fermionic part of the Lagrangian, ${\cal L}_f$, appear all the kinetic
terms for the fermion fields (gravitino, dilatino and gaugino).
Finally ${\cal L}_{int}$ contains the interactions among the various fields of the theory.

\subsection{$SU(3)$-structure manifolds}

\subsubsection{Generalities}

Calabi-Yau manifolds were proposed as internal spaces for
compactifications in view of the requirement that a four-dimensional
${\cal N}=1$ supersymmetry is preserved. Namely they admit a nowhere-vanishing,
globally defined spinor, which is covariantly constant with respect
to the (torsionless) Levi-Civita connection. However, there is a
wider class of manifolds for which the spinor is covariantly
constant with respect to a connection with torsion. These are called
manifolds with $SU(3)$-structure and clearly Calabi-Yau manifolds
belong in the class of $SU(3)$-structure manifolds.

More specifically, in order to define globally a nowhere-vanishing spinor on a
six-dimensional manifold one has to reduce the structure group
$SO(6)$ of the frame bundle. The simplest one can do is to reduce this group to
$SU(3)$, since then the decomposition of the spinor of $SO(6)$ reads
$\mathbf{4}=\mathbf{3}+\mathbf{1}$ and the spinor we are looking for
is the singlet, let us call it $\eta$. Then, we can use $\eta$ to
define the $SU(3)$-structure forms, which are a real two-form $J$ and a complex
three-form $\Omega$ defined as
\bea J_{mn} &=& \mp i\eta_{\pm}^{\dag}\gamma_{mn}\eta_{\pm}, \nn\\
    \Omega_{mnp} &=& \eta_-^{\dag}\gamma_{mnp}\eta_+,\nn\\
    \Omega^*_{mnp} &=& -\eta_+^{\dag}\gamma_{mnp}\eta_-, \eea where the signs denote the chirality of the spinor and the normalization is $\eta_{\pm}^{\dag}\eta_{\pm}=1$.
These forms are globally-defined and non-vanishing and they are
subject to the following compatibility conditions \bea J\w J\w J &=&
\frac{3}{4}i\Omega\w\Omega^*, \nn\\ J\w\Omega&=&0. \eea Moreover,
they are not closed forms but instead they satisfy
\bea\label{strfrms} dJ &=& \frac{3}{4}i({\cal
W}_1\Omega^*-{{\cal
W}}^*_1\Omega)+{\cal W}_4\w J+{\cal W}_3, \nn\\
    d\Omega &=& {\cal W}_1J\w J+{\cal W}_2\w J +{\cal W}_5^*\w
    \Omega. \eea
These expressions define the five intrinsic torsion classes, which
are a zero-form ${\cal W}_1$, a two-form ${\cal W}_2$, a three-form
${\cal W}_3$ and two one-forms ${\cal W}_4$ and ${\cal W}_5$. These
classes completely characterize the intrinsic torsion of the
manifold. Note that the classes ${\cal W}_1$ and ${\cal W}_2$ can be
decomposed in real and imaginary parts as ${\cal W}_1={\cal
W}_1^++{\cal W}_1^-$ and similarly for ${\cal W}_2$.

One can then classify the several types of manifolds in terms of the torsion classes.
We are not going to give an exhaustive list here (see \cite{Grana:2005jc} for more details), but it is worth noting
that in order for a manifold to be complex the classes ${\cal W}_1$ and ${\cal W}_2$ have to vanish and furthermore a K\"{a}hler manifold has
vanishing ${\cal W}_3$ and ${\cal W}_4$ as well. A Calabi-Yau manifold has
all the torsion classes equal to zero and the structure forms in this case are obviously closed.

\subsubsection{Homogeneous nearly-K\"{a}hler manifolds in six dimensions}

An interesting class of $SU(3)$-structure manifolds is that of
nearly-K\"{a}hler manifolds. In this case all the torsion classes
but ${\cal W}_1$ are vanishing. This suggests that the manifold is
not K\"{a}hler and not even complex.

The homogeneous nearly-K\"{a}hler manifolds in six dimensions have been classified in \cite{Butruille:2006} and they are the coset spaces
$\1,\2$ \footnote{Here we mean the non-symmetric coset space, obtained by the non-maximal embedding of $SU(2)\times U(1)$ in $Sp_4$. The maximal embedding yields a symmetric coset space, which does not admit an $SU(3)$-structure and is irrelevant for our purposes. Therefore we shall not use any special notation to distinguish these two coset spaces since we shall always refer to the non-symmetric one.} and $\3$ and the group manifold $SU(2)\times SU(2)$.
The first three cases are well-known to be the only non-symmetric coset spaces $S/R$ in six dimensions which preserve the rank, namely $rankS=rankR$. They have been studied extensively in \cite{Kapetanakis:1992hf} in the reduction of ten-dimensional gauge theories to four dimensions. Therefore it is interesting to study the reduction of the heterotic supergravity-Yang-Mills theory over these spaces and determine the corresponding effective actions in four dimensions.

A very interesting feature of the six-dimensional non-symmetric coset spaces is that they have simple and well-known geometry. Indeed, the most general $S$-invariant metric can be easily determined and the $S$-invariant $p$-forms are known explicitly. Let us mention here some general features of the geometric data of these spaces. A full account on these data can be found in Appendix B.

Concerning the most general $S$-invariant metric, it is always diagonal and depends on the number of radii
that each spaces admits. In particular $\1$ admits only one radius $R_1$, $\2$ admits two radii $R_1,R_2$ and $\3$ admits three radii $R_1,R_2,R_3$. Then the metric fluctuations can be parametrized by one, two and three scalar fields respectively.

All these spaces share the common feature that they do not admit $S$-invariant one-forms. On the contrary, $S$-invariant two-forms, which we shall denote by $\omega_i$, exist in all cases and in particular there is one for $\1$, two for $\2$ and three for $\3$. Moreover, all the three spaces admit two $S$-invariant three-forms, which we shall denote by $\rho_1$ and $\rho_2$. We collect the explicit expressions of these forms in Appendix B. Four-forms can also be found by dualizing the two-forms with respect to the six-dimensional Hodge star operator but they will not be useful in our framework.

An interesting fact about the invariant forms of the non-symmetric coset spaces is that they are intimately connected to the structure forms $J$ and $\Omega$, which specify the $SU(3)$-structure. As such, the knowledge of the $S$-invariant forms guarantees the knowledge of the $SU(3)$-structure and consequently of the intrinsic torsion classes. The real two form $J$ is a combination of the invariant two-forms $\omega_i$ and in particular
\bea J=R_1^2\omega_1 &&\mbox{for ~$\1$},\nn\\
        J=R_1^2\omega_1+R_2^2\omega_2 &&\mbox{for ~$\2$},\nn\\
        J= R_1^2\omega_1+R_2^2\omega_2+R_3^2\omega_3 &&\mbox{for ~$\3$}, \eea
where the different radii of the spaces appear in these expressions. On the other hand, the complex three-form $\Omega$ is always proportional to the combination $\rho_2+i\rho_1$ and particularly
\bea \Omega={R_1^3}(\rho_2+i\rho_1) &&\mbox{for ~$\1$},\nn\\
        \Omega={R_1^2R_2}(\rho_2+i\rho_1) &&\mbox{for ~$\2$},\nn\\
        \Omega= {R_1R_2R_3}(\rho_2+i\rho_1)&&\mbox{for ~$\3$}. \eea

The intrinsic torsion classes for each case appear in Appendix B. We note that in the case of $\1$ only ${\cal W}_1$ is non-vanishing and actually only its imaginary part. Therefore this manifold naturally admits a nearly-K\"{a}hler structure. In the other two cases, apart from ${\cal W}_1$ being non-vanishing, the ${\cal W}_2$ is generically different from zero as well. However, ${\cal W}_2$ also vanishes under the condition of equal radii ($R_1=R_2$ and $R_1=R_2=R_3$ respectively). It should be stressed that only when the latter condition holds the other two manifolds admit a nearly-K\"{a}hler structure too. Moreover this condition guarantees that the metric tensor is proportional to the Ricci tensor and therefore these manifolds become Einstein spaces \cite{MuellerHoissen:1987cq}.

\subsection{Coset Space Dimensional Reduction}

In the previous section we exhibited the fact that certain coset spaces admit a nearly-K\"{a}hler structure.
 Therefore we are naturally led to discuss the dimensional reduction over these spaces
 in the context of the Coset Space Dimensional Reduction (CSDR)
 \cite{Forgacs:1979zs},\cite{Kapetanakis:1992hf},\cite{Kubyshin:1989vd}
\footnote{For an interesting variant of this scheme see \cite{Lechtenfeld:2006wu}.} In the present section we present a brief reminder of the CSDR scheme. The basics of the
geometry of coset spaces are outlined in Appendix A.

Before describing the CSDR let us recall that the ansatz for the celebrated
Scherk--Schwarz reduction \cite{Scherk:1979zr} of a
higher-dimensional gauge field $\hat{A}$ on a group manifold $S$ has the form
\begin{equation}\label{SS}
\hat{A}=A_{\mu}dx^{\mu} + A_{I}(x)e^{I}(y),
\end{equation}
with $I=1,\ldots,{\rm dim} S$ and $e^{I}$ are the left-invariant
one-forms on the manifold. Then, this type of reduction on group manifolds amounts to
keeping only the $S_{L}$ singlets under the full isometry group
$S_{L} \times S_{R}$. This truncation can be described by the invariance condition,
\begin{equation}
{\cal L}_{X^{I}}\hat{A} = 0,
\end{equation}
with $X^{I}$ being the Killing vectors  dual to the right-invariant
one-forms\footnote{ Recall that the right-invariant vector fields
generate left translations.}. The Scherk--Schwarz reduction of the
metric is performed by enforcing a similar invariance condition
\begin{equation}\label{InvarianceCondition}
{\cal L}_{X^{I}}\hat{g}_{MN} =0.
\end{equation}The original CSDR of a multidimensional gauge field $\hat{A}$ on a
coset $B=S/R$ is a truncation described by a generalized invariance
condition
\begin{equation}\label{21}
{\cal L}_{X^{I}}\hat{A} = DW_{I}=dW_{I} + [\hat{A},W_{I}],
\end{equation}
where $W_{I}$ is a parameter of a gauge transformation associated
with the Killing vector $X_{I}$ of $S/R$. The relevant invariance
condition for the reduction of the metric is the same as in
(\ref{InvarianceCondition}),
namely the metric is considered invariant under the isometries of
the coset space. The generalized invariance condition (\ref{21})
together with the consistency condition
\begin{equation}\label{22}
\left[{\cal L}_{X^{I}}, {\cal L}_{X^{J}} \right] = {\cal
L}_{[X^{I},X^{J}]},
\end{equation}
impose constraints on the gauge field. The detailed analysis of the
constraints (\ref{21}) and (\ref{22}), given in refs.
\cite{Kapetanakis:1992hf},\cite{Forgacs:1979zs} provides us with the
four-dimensional unconstrained fields as well as with the gauge
invariance that remains in the theory after dimensional reduction.
Here we briefly state the results, which will be of considerable use
in the examples to follow after the general case.

\begin{itemize}
\item The four-dimensional gauge group $H$ is the centralizer of $R$ in
$G$\footnote{$G$ is the initial gauge group in higher dimensions, which in our cases will be identified with $E_8$.},
$H=C_G(R_G)$, provided that $R$ has an isomorphic image in $G$, $R_G$.
\item The representations of $H$ in which the four-dimensional
scalars\footnote{ Here we mean the internal components of the
multidimensional gauge field, which from the four-dimensional
viewpoint are Lorentz scalars.} transform can be determined by using the
decompositions
\begin{eqnarray}
S &\supset& R \nonumber \\
adjS &=& adjR+v
\end{eqnarray} and \begin{eqnarray}
G &\supset& R_{G} \times H \nonumber \\
 adjG &=&(adjR,1)+(1,adjH)+\sum(r_{i},h_{i}).
\end{eqnarray}

Then, if $v=\sum s_{i}$, where each $s_{i}$ is an irreducible
representation of $R$, there survives an $h_{i}$ multiplet for every
pair $(r_{i},s_{i})$, where $r_{i}$ and $s_{i}$ are identical
irreducible representations of $R$.
\item Finally, in order to determine how the four-dimensional spinor
fields transform we have to decompose the representation $F$ of the
initial gauge group, in which the fermions are assigned, under
$R_{G} \times H$, i.e.
\begin{equation}
F= \sum (t_{i},h_{i}),
\end{equation}
and the spinor of $SO(d)$ under $R$
\begin{equation}
\sigma_{d} = \sum \sigma_{j}.
\end{equation}
Here $d$ is the number of compactified dimensions. Then for each
pair $t_{i}$ and $\sigma_{i}$, where $t_{i}$ and $\sigma_{i}$ are
identical irreducible representations of $R$, an $h_{i}$ multiplet
of spinor fields survives in the four dimensional theory.
\end{itemize}

As another approach, we may use the following ansatz for the gauge
fields, which was shown in \cite{Chatzistavrakidis:2007by} to be
equivalent to the CSDR ansatz and it is similar to the
Scherk-Schwarz reduction ansatz:
\begin{equation}
\hat{A}^{\tilde{I}}(x,y) = A^{\tilde{I}}(x) +
\chi^{\tilde{I}}_{\alpha}(x,y)dy^{\alpha},
\end{equation}
where
\begin{equation}
\chi^{\tilde{I}}_{\alpha}(x,y) =
\phi^{\tilde{I}}_{A}(x)e^{A}_{\alpha}(y).
\end{equation} and $\tilde{I}$ is a gauge group index.
The objects $\phi_{A}(x)$, which take values in the Lie algebra of
$G$, are coordinate scalars in four dimensions and they can be
identified with Higgs fields. This procedure leads again to the CSDR
constraints, which in a compact form can be written as
\begin{equation}\label{constraints} D\phi^{\tilde{I}}_{i} =
F^{\tilde{I}}_{ai} = F^{\tilde{I}}_{ij} =0,
\end{equation}
where the index $i$ runs within the $R$ subgroup and $a$ is a coset
index (see Appendix A). These constraints will be used extensively
in the course of the reduction that will be performed in the
following sections.

\section{Reduction to four dimensions}

In the present section we focus on the bosonic part of the heterotic supergravity Lagrangian coupled to Yang-Mills
and perform its reduction from ten to four dimensions over the nearly-K\"{a}hler coset spaces S/R. Since the
K\"{a}hler potential and the superpotential of the four-dimensional theory can be obtained from
the bosonic part, this procedure will be sufficient to find the supergravity description in four
dimensions.

Let us also note that we shall work with dimensionless quantities in the intermediate stages of the procedure and we shall reinsert the dimensions in the next section where we shall deal with specific examples.

\subsection{Reduction of the metric and dilaton}

We begin by examining the ten-dimensional Einstein-Hilbert-dilaton
Lagrangian
\begin{equation}
\hat{e}^{-1}{\cal L} = -\frac{1}{2\hat{\kappa}^2}\hat{R} \hat{\ast}
{\bf 1}-\frac{1}{4\hat{\kappa}^2}d\hat{\phi}\w \hat{*}d\hat{\phi}.
\end{equation}
The general Kaluza-Klein ansatz for an $S$-invariant metric, including
all the fluctuations, would be
\begin{equation}\label{ouransatz}
d\hat{s}^2 = ds^2 + h_{\alpha\beta}(x,y)(dy^\alpha -
{\cal A}^\alpha(x,y))(dy^\beta - {\cal A}^\beta(x,y)),
\end{equation}
where $ds^2$ is the four-dimensional line element and ${\cal A}^\alpha$ denote the Kaluza-Klein gauge fields
\begin{equation}
{\cal A}^\alpha(x,y) = {\cal A}^I(x) K^{\alpha}_{(I)}(y), \ \ {\cal
A}^{I}(x) = {\cal A}^{I}_{\mu}(x) dx^{\mu}.
\end{equation}
Moreover $$ K_{(I)}(y) = K^{\alpha}_{(I)}(y) \frac{\partial}{\partial
y^{\alpha}},$$ are at most the $dim S + dim( N(R)/R)$ Killing
vectors of the coset $S/R$ or an appropriate subset\footnote{Recall that the maximal isometry group of a coset space $S/R$ is $S\times N(R)/R$. Here, $N(R)$ denotes the normalizer of $R$ in $S$, which
is defined as $N = \{ s \in S, \ \ \  sRs^{-1} \subset R \}.$ Note
that since $R$ is normal in $N(R)$ the quotient $N(R)/R$ is indeed a
group.}. However, an
additional constraint that coset reductions impose is that we cannot
allow Kaluza-Klein (KK) gauge fields from the maximal isometry group of
the coset $S/R$ to survive consistently \cite{Coquereaux:1986zf} -\cite{Chatzistavrakidis:2007pp}. In particular, tackling the consistency problem, direct calculations
lead to the result that when KK gauge fields take values in the
maximal isometry group of the coset space the lower-dimensional theory is, in general, inconsistent
with the original one. Full consistency of the effective Lagrangian
and field equations with the higher-dimensional theory is guaranteed
when the KK gauge fields are ($N(R)/R$)-valued. However, when the condition $rankS=rankR$ holds the group $N(R)/R$ is trivial. This is
the case for the spaces we consider and therefore the KK gauge
fields vanish due to the consistency requirement. Finally, the part of the internal metric $\gamma_{ab}(x)$ without the exponential has to be unimodular. Then the metric ansatz takes the form
\begin{equation}
d\hat{s}^2 = e^{2\alpha\varphi(x)} \eta_{mn} e^{m} e^{n} +
e^{2\beta\varphi(x)}\gamma_{ab}(x)e^{a}e^{b},
\end{equation} where $e^{2\alpha\varphi(x)}\eta_{mn}$ is the four-dimensional metric and $e^{2\beta\varphi(x)}\gamma_{ab}(x)$ is the internal metric, while $e^m$ are the one-forms of the orthonormal basis in four dimensions and $e^a$ are the left-invariant one-forms on the coset space.
In this ansatz we included exponentials which rescale the metric
components. This is always needed in order to obtain an action
without any prefactor for the four-dimensional Einstein-Hilbert part. In
order to fulfil this requirement we need to specify the values of $\alpha$ and $\beta$.

Following the standard procedure of reducing this action in the case
of a coset space (see e.g. \cite{Cvetic:2003jy}) and choosing
$\alpha = -\frac{\sqrt{3}}{4},\beta = -\frac{\alpha}{3}$, we find
that the reduced Lagrangian reads
\begin{eqnarray}\label{redlang1}
\mathcal{L}=-\frac{1}{2\kappa^2}\biggl(R*\mathbf{1}-P_{ab}\wedge*
P_{ab}+\frac{1}{2}d\varphi\wedge* d\varphi\biggl)-V_{grav},
\end{eqnarray}
with the potential $V_{grav}$ having the form
\begin{equation}\label{vgrav}
V_{grav}=\frac{1}{8\kappa^2}e^{2(\alpha-\beta)\varphi}(\gamma_{ab}\gamma^{cd}\gamma^{ef}f^{a}_{
\ ce}f^{b}_{ \
 df}+2\gamma^{ab}f^{c}_{ \ da}f^{d}_{ \ cb}+4\gamma^{ab}f_{iac}f^{ic}_{b}
 )*\mathbf{1},
\end{equation}
where the index $i$ runs in $R$. $\kappa$ is the gravitational coupling in four dimensions, related to the ten-dimensional one by $\kappa^2=\frac{\hat{\kappa}^2}{vol_6}$. In the expression (\ref{vgrav}) appear the structure constants of $S$ (see appendix A).

In the reduced Lagrangian the fields $P_{ab}$ are defined as
\beq P_{ab} = \frac{1}{2}\biggl[(\Phi^{-1})^{c}_{a}d\Phi^{b}_{c} +
(\Phi^{-1})^{c}_{b}d\Phi^{a}_{c}\biggl], \eeq
with $\Phi^a_b$
defined through the relation
\begin{equation} \gamma_{cd} =
\delta_{ab} \Phi^{a}_{c} \Phi^{b}_{d}.
\end{equation}
As such, $\Phi$ is a matrix of unit determinant,
generically containing scalar fields other than $\varphi$, and hence
there exists a set $(\Phi^{-1})^{b}_{a}$ of fields satisfying
\begin{equation}
(\Phi^{-1})^{c}_{a}(\Phi^{-1})^{d}_{b}\gamma_{cd} = \delta_{ab}.
\end{equation}
The corresponding kinetic term in (\ref{redlang1}) will provide the
kinetic terms for the extra scalars apart from $\varphi$, which are
generically needed to parametrize the most general $S$-invariant
metric and appear through the unimodular metric $\gamma_{ab}(x)$.
This concludes the reduction of the metric. On the other hand, the
dilaton is trivially reduced by $\hat{\phi}(x,y) = \phi(x)$, since
it is already a scalar in ten dimensions, leading to the term
$-\frac{1}{4\kappa^2} d\phi\w*d\phi$ in the reduced Lagrangian.

Let us now realize these results for the three spaces we examine. Concerning the fluctuations of the most general $S$-invariant metric $g_{ab}=e^{2\beta\vf}\gamma_{ab}$ we adopt the following parametrizations:
\bea\label{reparam}
\g_{ab}&=&\delta_{ab}~\mbox{for $\frac{G_2}{SU(3)}$}, \nn\\
\g_{ab}&=&\mbox{diag}(e^{2\gamma
\chi},e^{2\gamma \chi},e^{-4\gamma \chi},e^{-4\gamma
\chi},e^{2\gamma \chi},e^{2\gamma \chi})~\mbox{for $\frac{Sp_4}{SU(2)\times U(1)}$},\nn\\
\g_{ab}&=&\mbox{diag}(e^{2(\gamma\chi+\delta\psi)},e^{2(\gamma\chi
+\delta\psi)},e^{2(\gamma\chi-\delta\psi)},e^{2(\gamma\chi-\delta\psi)},e^{-4\gamma\chi},e^{-4\gamma
\chi})~\mbox{for $\frac{SU(3)}{U(1)\times U(1)}$},
\eea which clearly respect the unimodularity of $\g_{ab}$. In accordance with the expressions (\ref{reparam}) the metric fluctuations are parametrized by the scalar field $\vf(x)$ for $\1$, by the two scalar fields $\vf(x)$ and $\chi(x)$ for $\2$ and by the three scalar fields $\vf(x),\chi(x)$ and $\psi(x)$ for $\3$. Then, as far as the kinetic terms for the scalars are concerned, we immediately see that $P_{ab}=0$ in the first case, since there are no extra scalars apart from $\vf$. The situation changes in the other two cases. Indeed, for $\2$ we obtain the non-zero components for the fields $\Phi^a_b$
\bea \Phi^a_b =
\left\{\begin{array}{llc} e^{\gamma \chi}\delta^a_b, & a,b = 1,2,5,6
\\  e^{-2\gamma \chi}\delta^a_b, & a,b = 3,4 \\ \end{array}\right.
\eea
and consequently the corresponding ones for $P_{ab}$
\bea P_{ab} = \left\{\begin{array}{llc} \gamma d\chi\delta_{ab}, & a,b
= 1,2,5,6
\\  -2\gamma d\chi\delta_{ab}, & a,b = 3,4 \\ \end{array}\right.
\eea
Then, the kinetic term reads
\beq P_{ab}\w*P_{ab} = 12\gamma^2d\chi\w*d\chi, \eeq
namely the expected kinetic term for the scalar field $\chi$,
provided we make the choice $\gamma^2=\frac{1}{24}$.

In the same spirit, for the $\3$ case the metric provides us
with the fields
\bea \Phi^a_b = \left\{\begin{array}{llc} e^{
(\gamma\chi+\delta\psi)}\delta^a_b, & a,b = 1,2
\\  e^{ (\gamma\chi-\delta\psi)}\delta^a_b, & a,b = 3,4 \\ e^{-2\gamma  \chi}\delta^a_b, & a,b =5,6 \\ \end{array}\right.
\eea
from which we obtain
\bea P_{ab} = \left\{\begin{array}{llc} (\gamma d\chi+\delta
d\psi)\delta_{ab}, & a,b = 1,2
\\  (\gamma d\chi-\delta d\psi)\delta_{ab}, & a,b = 3,4 \\ -2\gamma d\chi\delta_{ab}, & a,b =5,6 \\ \end{array}\right.
\eea
Finally, this leads again to the expected kinetic terms for the
extra scalar fields
\beq P_{ab}\w*P_{ab} =
4(3\gamma^2d\chi\w*d\chi+\delta^2d\psi\w*d\psi), \eeq provided
again that we make the same choice for $\gamma$, while for $\delta$
we choose $\delta^2 = \frac{1}{8}$.

Having fixed the kinetic terms for all the scalar fields which parametrize the metric in each case we now turn to the four-dimensional potential. Exploiting the general expression (\ref{vgrav}) as well as the structure constants for each case (see appendix B), we determine the following potentials for the three spaces under consideration:
\bi
\item For $\1$~~\beq\label{potex1}  V_{grav} =-\frac{5}{\kappa^2}{e^{\frac{8\alpha}{3}\varphi}},\eeq
\item For $\2$~~\beq\label{potex2} V_{grav} =-\frac{1}{4\kappa^2}{e^{\frac{8\alpha}{3}\varphi}}(4e^{4\g\chi}+12e^{-2\g\chi}-e^{-8\g\chi}),\eeq
    \item For $\3$~~\beq\label{potex3} V_{grav} =-\frac{1}{4\kappa^2}{e^{\frac{8\alpha}{3}\varphi}}(6e^{4\g\chi}+6e^{-2(\g\chi-\delta\psi)}
        +6e^{-2(\g\chi+\delta\psi)}-e^{4(\g\chi+\delta\psi)}-e^{4(\g\chi-\delta\psi)}-e^{-8\g\chi}).\eeq
\ei

\subsection{Reduction of the gauge sector}

In this section we use the CSDR scheme to reduce the Yang-Mills part of the Lagrangian. The ansatz for the higher dimensional gauge field that solves the generalized
invariance condition (\ref{21}) is
\begin{equation}\label{apot}
\hat{A}^{\tilde{I}}=A^{\tilde{I}} + \phi^{\tilde{I}}_{A}e^{A},
\end{equation}
where $\tilde{I}$ is a gauge index and $A$ an $S$-index, which can
be split into indices $i,a$ running in the group $R$ and the coset respectively.
Calculating the field strength by
\begin{equation}
\hat{F} = \hat{d}\hat{A}^{\tilde{I}} + \frac{1}{2}f^{\tilde{I}}_{ \
\ \tilde{J} \tilde{K}} \hat{A}^{\tilde{J}} \wedge
\hat{A}^{\tilde{K}},
\end{equation}
we find that it can be written in terms of the four-dimensional
fields as
\begin{equation}\label{Pans}
\hat{F}^{\tilde{I}} = F^{\tilde{I}} + D\phi^{\tilde{I}}_{A} \wedge
e^{A} - \frac{1}{2}F^{\tilde{I}}_{AB} e^{A} \wedge e^{B}.
\end{equation}
In the last expression
\begin{equation} F^{\tilde{I}} = dA^{\tilde{I}} +
\frac{1}{2}f^{\tilde{I}}_{ \ \ \tilde{J} \tilde{K}} A^{\tilde{J}}
\wedge A^{\tilde{K}}
\end{equation}
is the four-dimensional gauge field strength, while its internal
components have the form
\begin{equation}
F^{\tilde{I}}_{AB} = f^{C}_{ \ \ AB}\phi^{\tilde{I}}_{C}-
f^{\tilde{I}}_{ \ \ \tilde{J} \tilde{K}}\phi^{\tilde{J}}_{A}
\phi^{\tilde{K}}_{B}.
\end{equation}
Finally
\begin{equation}
D\phi^{\tilde{I}}_{A}= d\phi^{\tilde{I}}_{A} +  f^{\tilde{I}}_{ \ \
\tilde{J} \tilde{K}}A^{J} \phi^{\tilde{K}}_{A}
\end{equation}
is the covariant derivative of the internal components of the gauge
field.

To reduce the higher dimensional Yang-Mills Lagrangian we dualize
eq.~(\ref{Pans})
\begin{equation}\label{DualPans}
\hat{\ast}\hat{F}^{\tilde{I}} =  \ast_{4}F^{\tilde{I}}\wedge
vol_{6} + e^{\alpha\varphi-\beta\varphi}
\ast_{4}D\phi_{A}^{\tilde{I}} \wedge \ast_{6}\tilde{e}^{A} -
\frac{1}{2} e^{2\alpha\varphi-2\beta\varphi} F^{\tilde{I}}_{AB}
vol_{4} \wedge \ast_{6}(\tilde{e}^{A} \wedge \tilde{e}^{B}),
\end{equation}
where we have defined
\begin{equation}
\tilde{e}^{a} = (\Phi^{-1})^{a}_{b}e^{b}, \ \ \ \tilde{e}^{i} =
e^{i}_{a}(\Phi^{-1})^{a}_{b}e^{b}.
\end{equation}
Inserting the expressions (\ref{Pans}) and (\ref{DualPans}) in the corresponding term in the Lagrangian
$$ {\cal L}_{gauge} = -\frac{\alpha'}{4\hat{\kappa}^2}\hat{e}e^{-\miso\hat{\phi}} Tr \hat{F} \wedge \hat{\ast}\hat{F}$$ and using that the determinant of the metric is $\hat{e}=e^{2\alpha\varphi}$ we obtain
\begin{eqnarray} \label{KKYM}
{\cal L}_{gauge} &=& -\frac{\alpha'}{4\kappa^2} e^{-\miso\phi}\biggl[e^{-2\alpha\varphi}F^{\tilde{I}}
\wedge \ast_{4}F^{\tilde{I}} \wedge vol_{6} + e^{-2\beta\varphi}
D\phi^{\tilde{I}}_{A} \wedge \ast_{4}D\phi_{B}^{\tilde{I}} \wedge
{e}^{A} \wedge \ast_{6} \tilde{e}^{B} \nonumber
\\ &+& \frac{1}{4}e^{2\alpha\varphi-4\beta\varphi} F_{AB}F_{CD} vol_{4} \wedge {e}^{A} \wedge
{e}^{B} \wedge \ast_{6}(\tilde{e}^{C} \wedge \tilde{e}^{D})\biggl],
\end{eqnarray}
To reduce eq.~(\ref{KKYM}) we must impose the CSDR constraints
\begin{equation}\label{newcon1}
D\phi^{\tilde{I}}_{i} =0, \ \ F^{\tilde{I}}_{ij} =
F^{\tilde{I}}_{aj} = 0.
\end{equation}
Collecting the various terms we obtain the Lagrangian
\begin{equation}\label{gaugel}
{\cal L}_{gauge}
=-\frac{\alpha'}{4\kappa^2}e^{-\miso\phi}\biggl[e^{-2\alpha\varphi}F^{\tilde{I}} \wedge
\ast_{4} F^{\tilde{I}} \wedge vol_{6} +e^{-2\beta\varphi}\gamma^{{a}
{b}} D\phi_{{a}}^{\tilde{I}} \wedge \ast_{4}D\phi_{{b}}^{\tilde{I}}
\wedge vol_{6}\biggl]-V_{gauge},
\end{equation}
consisting of the kinetic term for the four-dimensional gauge
fields, the kinetic term for the coordinate scalars, which will be
identified with the Higgs fields, and a scalar potential of the form
\begin{equation}\label{vgauge}
V_{gauge} = \frac{\alpha'}{8\kappa^2}{e^{-\miso\phi+2(\alpha-2\beta)\varphi}}
\gamma^{{a} {c}} \gamma^{{b} {d}} F_{{a} {b}}F_{{c} {d}}.
\end{equation}

\subsection{Reduction of the three-form}

Our next task is to perform the reduction of the term in the bosonic
Lagrangian containing the three-form field strength $\hat{H}_{(3)}$,
\begin{equation}\label{L3}
{\cal L}_{H} = -\frac{1}{4\hat{\kappa}^2}\hat{e}
e^{-\hat{\phi}}\hat{H}_{(3)} \wedge \hat{*} \hat{H}_{(3)}.
\end{equation}
The three-form $\hat{H}${\footnote{We shall omit the subscript (3)
from now on to avoid having too heavy notation.} is given in
general by \beq\label{hexp} \hat{H} =  \hat{d}\hat{B} -\frac{\alpha'}{2}(
\hat{\omega}_{YM}-\hat{\omega}_L). \eeq Here $\hat{B}$ is the abelian
two-form potential, which we expand in the $S$-invariant forms of the internal space as
\begin{equation}\label{twoform}
\hat{B}= B(x) + b^i(x)\omega_i(y).
\end{equation}
The expansion forms $\omega_i(y)$ are the $S$-invariant two-forms on the
internal space.
Note that a possible term of the form $B_{a}\wedge e^a$ in the expansion of the $B$-field is forbidden since $S$-invariant one-forms do not exist on the spaces we consider.

The $\hat{\omega}_{YM}$ in eq. (\ref{hexp}) is the usual Yang-Mills-Chern-Simons form,
\beq \hat{\omega}_{YM} =
Tr(\hat{F}\wedge\hat{A}-\frac{1}{3}\hat{A}\wedge\hat{A}\wedge\hat{A}),
\eeq
and the $\hat{\omega}_L$ is the Lorentz-Chern-Simons form, constructed from
the modified spin connection $\hat{\tilde{\theta}}$,
\beq \hat{\omega}_L=Tr(\hat{\tilde{\theta}}\wedge
d\hat{\tilde{\theta}}+\frac{2}{3}\hat{\tilde{\theta}}\w\hat{\tilde{\theta}}\w\hat{\tilde{\theta}}),\eeq where the traces are evaluated in the
adjoint representation of the gauge group and the vector representation of the Lorentz group respectively. The modified spin connection is given in terms of the Levi-Civita one by
$\hat{\tilde{\theta}}=\hat{\theta}-\miso H. $ These two
corrections are necessary to cancel completely the anomalies (gauge,
gravitational and mixed) of ${\cal N}=1$, $D=10$ supergravity
coupled to Yang-Mills. The Bianchi identity associated with the three-form reads
\beq d\hat{H}=\frac{\alpha'}{2} (Tr(\hat{\tilde{R}}\w\hat{\tilde{R}})-Tr(\hat{F}\w\hat{F})), \eeq where again $\hat{\tilde{R}}$ is calculated using the modified spin connection. Note that in the following we shall not reduce the
Lorentz-Chern-Simons form because it is not needed in the minimal supergravity Lagrangian.

Differentiating eq.~(\ref{twoform}) we obtain
\beq\label{threeform} \hat{H}^{(B)} \equiv \hat{d}\hat{B} = dB +
(db^i)\w\omega_i +b^id\omega_i. \eeq
Let us note here that unlike
the case of CY compactifications, where the expansion forms are
harmonic and hence closed, here we expand in forms that are not
closed and thus an extra term appears in eq. (\ref{threeform}).
Moreover, using the eqs. (\ref{apot}) and (\ref{Pans})
and imposing the CSDR constraints, we find\footnote{We hereby adopt the usual notation $e^a\w e^b\equiv e^{ab}$, $e^a\w e^b\w e^c \equiv e^{abc}$ in order to avoid using repeatedly the symbol of the wedge product.}  \bea
\hat{\omega}_{YM} &=& \omega_{YM} + Tr(\phi_bD\phi_a)\w e^{ab}-\miso
Tr(\phi_cF_{ab})e^{abc}-\frac{1}{3}Tr(\phi_a\phi_b\phi_c)e^{abc},
 \eea
where \beq \omega_{YM} = Tr(F\w A)-\frac{1}{3}Tr(A\w A\w A) \eeq is
the four-dimensional Yang-Mills-Chern-Simons form. Then the full expression for the field strength becomes \bea \hat{H}\label{hhat}
&=& dB + db^i\w\omega_i+ \frac{\alpha'}{2}Tr(\phi_aD\phi_b)\w e^{ab} \nn \\
 &+&b^id\omega_i- \frac{\alpha'}{3}Tr(\phi_a\phi_b\phi_c)e^{abc}+\frac{\alpha'}{4}
f_{ab}^dTr(\phi_c\phi_d)e^{abc}, \eea where we have used the fact that
the form of the internal gauge field strength is \beq F_{ab} =
f_{ab}^C\phi_C-[\phi_a,\phi_b]. \eeq

To write down the dimensionally reduced form of the Lagrangian
(\ref{L3}) we dualize the three-form $\hat{H}$ with respect to the ten-dimensional Hodge star operator,
\bea \hat{*}\hat{H} &=&
e^{-6\alpha\varphi}*_4dB\w vol_6
+ e^{-2\alpha\varphi-4\beta\varphi}*_4db^i\w*_6\omega_i\nn\\&+&\frac{\alpha'}{2} e^{-2\alpha\varphi-4\beta\varphi}Tr(\phi_a*_4D\phi_b)\w*_6\tilde{e}^{ab} \nn \\
&+&e^{-6\beta\varphi}b^ivol_4\w\ast_6 d\omega_1-
\frac{\alpha'}{3}e^{-6\beta\varphi}Tr(\phi_a\phi_b\phi_c)vol_4\w*_6\tilde{e}^{abc}
\nn
\\&+&\frac{\alpha'}{4} e^{-6\beta\varphi}f_{ab}^dTr(\phi_c\phi_d)vol_4\w*_6\tilde{e}^{abc}. \eea

Substituting these expressions into the Lagrangian (\ref{L3}),
we initially obtain\footnote{We shall omit in the following expressions the subscripts of the star operator since it is obvious whether it corresponds to $\ast_4$ or $\ast_6$.}
\begin{eqnarray}\label{RL3}
{\cal L}_{H} = -\frac{1}{4{\kappa}^2}
e^{-\phi}\biggl[&&e^{-4\alpha\varphi}dB\w
                    *dB\w vol_6\nn\\
                    &+&e^{-4\beta\varphi}db^i\w *db^j\w \omega_i\w*\omega_j\nn \\
            &+&  \frac{\alpha'^2}{4}e^{-4\beta\varphi}Tr(\phi_aD\phi_b)\w
                    Tr(\phi_{c}*D\phi_{d})\w e^{ab}\w*\tilde{e}^{cd}
                    \nn\\&+&\alpha' e^{-4\beta\varphi}db^i\w Tr(\phi_a*D\phi_b)\w \omega_i\w\ast\tilde{e}^{ab}  \nn \\
            &+&e^{2\alpha\varphi-6\beta\varphi}b^ib^jvol_4\w d\omega_i\w\ast d\omega_j\nn\\&+& \frac{1}{9}\alpha'^2e^{2\alpha\varphi
            -6\beta\varphi}Tr(\phi_{a}\phi_b\phi_{c})
            Tr(\phi_{d}\phi_e\phi_{f})vol_4\w e^{abc}\w*\tilde{e}^{def} \nn \\
            &-& \frac{2\alpha'}{3}e^{2\alpha\varphi
            -6\beta\varphi}b^iTr(\phi_a\phi_b\phi_c)vol_4\w d\omega_i\w\ast\tilde{e}^{abc}\nn\\&+&\frac{\alpha'}{2}e^{2\alpha\varphi
            -6\beta\varphi}b^iTr(f_{ab}^d\phi_c\phi_d)vol_4\w d\omega_i\w\ast\tilde{e}^{abc}
            \nn\\&+&
            \frac{\alpha'^2}{16}e^{2\alpha\varphi
            -6\beta\varphi}Tr(f_{ab}^d\phi_c\phi_d)Tr(f_{ef}^h\phi_g\phi_h)vol_4\w e^{abc}\w*\tilde{e}^{efg}\nn\\
            &-&\frac{\alpha'^2}{6}e^{2\alpha\varphi
            -6\beta\varphi}Tr(\phi_a\phi_b\phi_c)Tr(f_{de}^g\phi_f\phi_g)vol_4\w
            e^{abc}\w*\tilde{e}^{def}\biggl].
\end{eqnarray} Obviously all terms are proportional to $vol_6$, the volume of the
internal space, albeit not all of them in a certain manner. There
appear the combinations\footnote{The $\Phi$s appearing through $\tilde{e}^a$ only contribute extra exponentials so for the moment we can ignore the tildes in this discussion.} $\omega_i\w*\omega_j$, $\omega_i\w*e^{ab}$, $d\omega_i\w\ast d\omega_j$, $d\omega_i\w\ast e^{abc}$, $e^{ab}\w*e^{cd}$ and $e^{abc}\w*e^{def}$. The last two are
completely determined by certain identities which are presented in appendix A. Concerning the first four, they depend on the geometric
data of the spaces we are going to use and they are actually related to the nearly-K\"{a}hler structure as we shall see in the examples to be presented in the following section. In order to keep track of the
general case we define
\bea\label{data} \omega_i\w*\omega_j &=& m \delta _{ij}vol_6, \nn\\
        \omega_i\w*e^{ab} &=& \epsilon_i^{ab}vol_6. \nn\\
        d\omega_i\w*d\omega_j &=&
            (n_1\delta_{ij}+n_2\epsilon_{ij})vol_6, \nn \\
        d\omega_i\w*e^{abc} &=& \epsilon_i^{abc}vol_6.
        \eea
The constants $m,n_1,n_2$ are fixed numbers which can be easily determined for each homogeneous nearly-K\"{a}hler manifold and their specific values can be found in Appendix B, along with the details related to
the $\epsilon$-symbols used here.

Finally, the usual duality transformation on $B$ \beq e^{4\alpha\varphi}d\theta=\ast dB \eeq provides a pseudoscalar $\theta$, which moreover has an axionic coupling to the gauge field strength as shown in \cite{Derendinger:1985kk}.

After this preparation we are ready to write down the final form of
the general Lagrangian, which is
\begin{eqnarray}\label{RL4}
{\cal L}_{H} = -\frac{1}{4\kappa^2}
e^{-\phi}\biggl[&&e^{4\alpha\varphi}d\theta\w*d\theta
-e^{4\alpha\varphi}\theta F^I\w{F}^I+m
e^{-4\beta\varphi}db^i\w
*db^i\nn\\&+&  \alpha'e^{-4\beta\varphi}\epsilon_i^{ab}db^i\w Tr(\phi_a*D\phi_b)\nn\\&+&\frac{\alpha'^2}{4}e^{-4\beta\varphi}Tr(\phi_a\olra{D}\phi_b)\w
                    Tr(\phi_{a}*\olra{D}\phi_{b})\biggl]\w vol_6\nn\\&-&V_{H}\w vol_6, \eea
with the potential appearing in this Lagrangian having the general form
 \bea\label{hpotentialgen}  V_{H}=\frac{1}{4\kappa^2}
        \frac{e^{-\phi+4\alpha\varphi}}{R^6}\biggl[&&b^ib^j(n_1\delta_{ij}+n_2\epsilon_{ij}) - \frac{2\alpha'}{3}\epsilon^{abc}_ib^iTr(\phi_a\phi_b\phi_c)\nn\\&+&\frac{\alpha'}{2}\epsilon^{abc}_ib^iTr(f_{ab}^d\phi_c\phi_d)
            +\frac{2\alpha'^2}{3}Tr(\phi_{a}\phi_b\phi_{c})^2\nn\\&+&
            \frac{\alpha'^2}{16}Tr(f_{ab}^d\phi_{c}\phi_d)Tr(f_{[ab}^d\phi_{c]}\phi_d)
            \nn \\
            &-&\alpha'^2Tr(\phi_a\phi_b\phi_c)Tr(f_{ab}^d\phi_c\phi_d)\biggl]vol_4,
\end{eqnarray}
where the contractions are performed with the inverse unimodular metric
$\gamma^{ab}$ \footnote{Due to this fact there will appear extra
exponentials in the several terms when we shall deal with specific
examples.} and we have reinserted R to keep track of the dimensions. Consequently the
scalar potential obtained from the reduction of the metric and gauge
field on non-symmetric coset spaces is further modified. In the following section we shall
focus on specific examples where the invariant forms are
explicitly known and the potential can be brought in a more transparent
form.

\section{Applications}

In this section we provide realizations of the general case that
has been presented so far, treating in detail the three specific examples
of internal manifolds we discussed before. The gravitational sector of these models has been already treated and the results appear in section 3.1. In particular, the relevant potentials are given by the expressions (\ref{potex1}}), (\ref{potex2}) and (\ref{potex3}) respectively. However, we shall reexamine these potentials in order to be able to obtain the nearly-K\"{a}hler limit in all cases. Indeed, we have already discussed that only the manifold $\1$ is genuinely a nearly-K\"{a}hler manifold, while the manifolds $\2$ and $\3$ become nearly-K\"{a}hler only under the condition of equal radii. However, in the naive parametrizations (\ref{reparam}) for the unimodular metric $\gamma_{ab}$ it is far from obvious how one can take the nearly-K\"{a}hler limit. In order to discuss this issue we shall see that it is convenient to perform appropriate redefinitions of the scalar moduli fields which appear in the four-dimensional theory. These redefinitions will be carefully done in order to preserve the correctly normalized kinetic terms for the corresponding fields. In other words after the redefinitions no mixed terms will appear. Moreover in the present section we proceed to the evaluation of the potentials arising from the gauge and three-form sectors.

\subsection{Example based on ${\1}$}

\

{\it{Gravity sector}}

\

As we have already discussed the $\1$ is a genuine nearly-K\"{a}hler manifold. However, it is convenient to perform the following redefinition of the scalar fields $\phi$ and $\vf$,
\bea\label{redef1n}
        \t{\phi}&=&\miso(-\phi-4\alpha\varphi), \nn\\
        \t{\varphi}&=&\miso(-\varphi+4\alpha\phi). \eea
Using this redefinition the potential (\ref{potex1}) in terms of the redefined scalar fields takes the form, \beq\label{potex1r}  V_{grav} =-\frac{15}{\kappa^2}\frac{e^{-\t{\phi}}}{R_1^2},\eeq with the radius $R_1(x)$ given by the relation \beq\label{radius1} R_1^2(x)=\frac{R^2}{3}e^{-\frac{\t{\vf}}{\sqrt{3}}}, \eeq where we have now reinserted the dimensions. Let us note that the exponential involving the dilaton $\t{\phi}$ in the expression (\ref{potex1r}) appears because we are using the metric in the Einstein frame. As it was noted in \cite{Benmachiche:2008ma} the correct four-dimensional field variables arise in the string frame. The transition from the Einstein frame to the string frame is done by multiplying with a dilaton-dependent factor. In our conventions this transition amounts to multiplying the fields by a factor $e^{\t{\phi}/2}$. The same remark will also apply in the following two cases.

\

{\it{Gauge sector}}

\

The four-dimensional gauge sector has been separately treated in detail in \cite{Manousselis:2000aj}. Here we review the basic steps of the computation of the potential since they will also be useful in determining the potential arising from the three-form sector.

According to the general rules of the CSDR presented in section 2.3, we have
to decompose the adjoint representation of $G=E_8$ in
representations of $R=SU(3)$. Hence, we use the decomposition
\begin{eqnarray}\label{dec1}
E_{8} &\supset& SU(3) \times E_{6}\nonumber \\ 248 &=& (8,1) +
(1,78) + (3,27) + (\overline{3},\overline{27})
\end{eqnarray}
and we choose $SU(3)$ to be identified with $R$. The $SU(3)$
content  of the $\1$ vector and spinor is $ 3 + \overline{3}$
and $1+3$ as can be read from tables 1 and 2 of Appendix B. The resulting
four-dimensional gauge group is $ H = C_{E_{8}}(SU(3)) = E_{6}$,
which contains fermion and complex scalar fields transforming as 78, 27 and 27 respectively,
according to the rules stated in section 2.3. The number of fermionic generations surviving in four
dimensions is one. This agrees with
the general result, based on the Atiyah-Singer index theorem, that
the fermion families are equal to half of the Euler characteristic of
the internal space. Indeed, $\1$ has
Euler characteristic 2.

In order to determine the potential the
decomposition of the adjoint of the specific $S$ under $R$ has to be examined, i.e.
\begin{eqnarray}\label{dec2}
G_{2} &\supset& SU(3) \nonumber \\ 14 &=& 8+3+\overline{3}.
\end{eqnarray}
Corresponding to this decomposition we introduce the generators of
$G_{2}$
\begin{equation}
Q_{G_{2}} = \{ Q^{a},Q^{\rho},Q_{\rho}\},
\end{equation}
where $a=1,\ldots,8$ correspond  to the $8$ of $SU(3)$, while $\rho
= 1,2,3$ correspond to $3$ and $\overline{3}$.
The potential of any theory reduced over $G_{2}/SU(3)$ can be
written in terms of the fields
\begin{equation}\label{ared}
\{ \phi_{a} , \phi^{\rho} ,\phi_{\rho} \},
\end{equation}
which correspond to the decomposition (\ref{dec2}) of $G_{2}$ under
$SU(3)$. The $\phi_{a}$ are equal to the generators of the  $R$
subgroup.
The generators of $E_{8}$ should also be divided according to the embedding (\ref{dec1}),
\begin{equation}
Q_{E_{8}} = \{ Q^{a} , Q^{\alpha},Q^{i\rho},Q_{i\rho} \}
\end{equation}
with $a = 1,\ldots,8$,  $\alpha = 1,\ldots,78$, $i=1,\ldots,27$,
$\rho=1,2,3$.

In order to express the resulting four-dimensional potential in terms of the
unconstrained scalar fields, let us call them $\beta$, the solutions of the constraints (\ref{constraints}) have to be specified. In terms of the genuine Higgs fields these solutions are \cite{Manousselis:2000aj}
\begin{equation}
\phi^{a}=Q^{a},\  \phi_{\rho}=R_1\beta^{i}Q_{i\rho},\
\phi^{\rho}=R_1\beta_{i}Q^{i\rho}.
\end{equation}
In turn the Higgs potential can be expressed in terms of the genuine
Higgs fields $\beta^i$ and the result is \bea\label{pot44} V_{gauge}(\beta)&=&\frac{\alpha'}{8\kappa^2}
e^{-\miso\t{\phi}}\biggl(\frac{{8}}{R_1^4}-
\frac{40}{3R_1^2}\beta^{2} -
\left[\frac{{4}}{R_1}d_{ijk}\beta^{i}\beta^{j}\beta^{k} + h.c
\right] \nn\\&+&\beta^{i}\beta^{j}d_{ijk}d^{klm}\beta_{l}\beta_{m}+
\frac{11}{4}\sum_{\alpha}\beta^{i}(G^{\alpha})_{i}^{j}
\beta_{j}\beta^{k}(G^{\alpha})_{k}^{l}\beta_{l}\biggl)vol_4, \eea where
$d^{ijk}$ is the symmetric invariant $E_{6}$ tensor, and
$(G^{\alpha})^{i}_{j}$ are defined as in \cite{Kephart:1981gf}. Here and in the ensuing expressions we use the notation $\beta^2=\beta_i\beta^i$. We note that in order to write down the expression (\ref{pot44}) for the four-dimensional potential we also used the redefinition (\ref{redef1n}) and the expression (\ref{radius1}) for the radius $R_1$. We observe that the exponential prefactor after the redefinition is $e^{-\miso\t{\phi}}$, a welcome result since this is the four-dimensional dilaton.

As it was argued in \cite{Manousselis:2000aj}, in the potential (\ref{pot44}) we can read directly the $F$-terms and $D$-terms. We shall see in the following section that the $F$-terms can be obtained from a superpotential. Moreover this potential contains terms which in a Minkowskian four-dimensional theory could be interpreted as soft scalar masses and soft trilinear terms.

\

{\it{Three-form~sector}}

\

Next let us work out the contribution of the three-form field
strength. Here it is important to know the invariant forms on the
coset space, which can be found in the Appendix B.

The Lagrangian (\ref{RL4}) can now be written as
\begin{eqnarray}\label{RL5ex1}
{\cal L} = -\frac{1}{4\kappa^2}
e^{-{\phi}}\biggl[&&e^{4\alpha\varphi}d\theta\w*d\theta-e^{4\alpha\varphi}\theta F^I\w F^I
            + 3e^{-4\beta\varphi}db\w *db\nn\\ &&+  \alpha'e^{-4\beta\varphi}\epsilon_1^{ab}db\w Tr(\phi_a*D\phi_b) \nn \\
            &&+ \frac{\alpha'^2}{4}e^{-4\beta\varphi}Tr(\phi_a\olra{D}\phi_b)\w
                    Tr(\phi_{a}*\olra{D}\phi_{b})\biggl]\w vol_6\nn\\&&-V_{H}\w vol_6,
\end{eqnarray}
since there is only one $G_2$-invariant two-form ($i=1$) and therefore one scalar $b^1\equiv \frac{b}{\sqrt{3}}$ arising from the internal components of the $B$-field. We have also substituted the value $m=3$ of the constant $m$.
The potential term, which is given in general in eq. (\ref{hpotentialgen}), takes the form
\bea V_{H}=\frac{1}{4\kappa^2}
        {{e^{-\t{\phi}}}}\biggl[&&{12}(b^1)^2  - \frac{2\alpha'}{3}\epsilon^{abc}_1b^1Tr(\phi_a\phi_b\phi_c)\nn\\&&+\frac{\alpha'}{2}\epsilon^{abc}_1b^1Tr(f_{ab}^d\phi_c\phi_d)\nn\\
            &&+\frac{2\alpha'^2}{3}Tr(\phi_{a}\phi_b\phi_{c})^2\nn\\&&+
            \frac{\alpha'^2}{16}Tr(f_{ab}^d\phi_{c}\phi_d)Tr(f_{[ab}^d\phi_{c]}\phi_d)
            \nn \\
            &&-\alpha'^2Tr(\phi_a\phi_b\phi_c)Tr(f_{ab}^d\phi_c\phi_d)\biggl]vol_4,
\end{eqnarray}
where we have substituted the value of the constant $n_1=12$ ($n_2$ is irrelevant in this case since the corresponding term is absent). We observe that after the redefinitions (\ref{redef1n}) the exponential prefactor for the three-form potential takes the welcome form $e^{-\t{\phi}}$. As before we would like to express the potential in terms of the genuine
Higgs fields by using the same complex scalars we defined for the
gauge sector in eq. (\ref{ared}). Thus we obtain the result
\begin{eqnarray}\label{hpotex1}
V_{H} = \frac{1}{\kappa^2}
e^{-\t{\phi}}\biggl[&&\frac{b^2}{R_1^6}+\frac{\sqrt{2}}{R_1^3}i\alpha'b(d_{ijk}\beta^i\beta^j\beta^k-h.c.)
            +2\alpha'^2\beta^i\beta^j\beta^kd_{ijk}d^{lmn}\beta_l\beta_m\beta_n\nn\\
            &&+\frac{3}{R_1^2}\alpha'^2(\beta^2)^2-\frac{\sqrt{6}}{R_1}\alpha'^2\beta^2(d_{ijk}\beta^i\beta^j\beta^k+h.c.)\biggl]vol_4.
\eea

\

\subsection{Example based on ${\2}$}

\

{\it{Gravity sector and the nearly-K\"{a}hler limit}}

\

We have already mentioned that the manifold $\2$ is a nearly-K\"{a}hler manifold only under the condition of equal radii. In order to clarify how this nearly-K\"{a}hler limit can be obtained we perform the following redefinition of the moduli fields $\phi,\vf$ and $\chi$, \bea\label{redef2n} \t{\phi} &=&
        -\miso(\phi+4\alpha\varphi),\nn\\
        \t{\varphi}&=&-\frac{\sqrt{2}}{2}(\phi-\frac{4\alpha}{3}\varphi+4\gamma\chi),\nn\\
        \t{\chi}&=&-\miso(\phi-\frac{4\alpha}{3}\varphi-8\gamma\chi).\eea
Using this redefinition we can show that the potential (\ref{potex2}) can be written in terms of the redefined scalar fields in the form \beq\label{potex2r} V_{grav} =-\frac{1}{4\kappa^2}{e^{-\t{\phi}}}(\frac{4}{R_2^2}+\frac{12}{R_1^2}-\frac{R_2^2}{R_1^4}),\eeq where the radii $R_1(x)$ and $R_2(x)$ of $\2$ are given by \bea\label{radii2} R_1^2&=&R^2e^{-\frac{\t{\vf}}{\sqrt{2}}}, \nn\\
        R_2^2&=&R^2e^{-\t{\chi}}.
\eea
Then it is relatively straightforward how the nearly-K\"{a}hler limit can be obtained. In the case of $\2$ the nearly-K\"{a}hler limit is obtained when $R_1=R_2$ i.e. when $\frac{\t{\vf}}{\sqrt{2}}=\t{\chi}$. Then the expression (\ref{potex2r}) takes the limiting form,
\beq\label{potlimit} V_{grav} =-\frac{15e^{-\t{\phi}}}{4\kappa^2}\frac{1}{R_1^2}, \eeq which looks formally the same as the one we obtained in the case of $\1$.

\

{\it{Gauge sector}}

\

As far as the gauge sector is concerned, in the same spirit as in the previous case, in the present one we have to specify the decompositions which are relevant for our analysis, $$E_{8} \supset
SU(3) \times E_{6} \supset SU(2) \times U(1) \times E_{6}.$$ The
decomposition of $248$ of $E_{8}$ under $SU(3) \times E_{6}$ was
given in the previous example, while under $ (SU(2) \times U(1))
\times E_{6}$ it is the following,
\begin{eqnarray}\label{dec4}
248 =
(3,1)_{0}+(1,1)_{0}+(1,78)_{0}+(2,1)_{3}+(2,1)_{-3}\nonumber\\
+(1,27)_{-2}
+(1,\overline{27})_{2}+(2,27)_{1}+(2,\overline{27})_{-1}.
\end{eqnarray}
In the present case $R$ is chosen to be identified with the $SU(2)
\times U(1)$ of the latter of the above decompositions. Therefore
the resulting four-dimensional gauge group is $H=C_{E_{8}}(SU(2)
\times U(1))= E_{6} \times U(1)$.

Concerning the abelian factor which appears in the four-dimensional gauge group $H$, we note that the corresponding gauge boson surviving in four dimensions becomes massive at the compactification scale \cite{Witten:1984dg} and therefore it does not contribute in the anomalies; it corresponds only to a global symmetry.

In order to proceed in our analysis, keeping in mind our latter remark, according to tables 3 and 4 the $R=SU(2) \times U(1)$ content of $\2$ vector and spinor are $1_{2}+1_{-2}+2_{1}+2_{-1}$ and
$1_{0}+1_{-2}+2_{1}$ respectively. Thus, applying the CSDR rules we
find that the surviving fields in four dimensions can be organized
in a ${\cal N}=1$ vector supermultiplet $V^{\alpha}$ which
transforms as $78$ under $E_{6}$ and two chiral supermultiplets ($B^{i}$,
$\Gamma^{i}$), transforming as $27$ under $E_{6}$. The number of fermion generations for this model is two, as
expected since the Euler characteristic of this space is four.

To determine the potential the decomposition of the adjoint of the specific $S$ under
$R$ has to be examined further, i.e. $$Sp(4) \supset (SU(2) \times U(1))_{non-max.}$$
\begin{equation}\label{dec3}
10 = 3_{0}+1_{0}+1_{2}+1_{-2}+2_{1}+2_{-1}.
\end{equation}
Then, according to the latter decomposition, the generators of
$Sp(4)$ can be grouped as follows,
\begin{equation}
Q_{Sp(4)} = \{ Q^{\rho},Q,Q_{+},Q^{+},Q^{a},Q_{a}\},
\end{equation}
where $\rho$ takes values $1,2,3$ and $a$ takes the values $1,2$.
Furthermore the decomposition
(\ref{dec3}) suggests the following change in the notation of the
scalar fields
\begin{equation}\label{redefex2}
\{ \phi_{I}, I=1,\ldots,10\} \longrightarrow ( \phi^{\rho}, \phi,
\phi_{+}, \phi^{+}, \phi^{a}, \phi_{a}),
\end{equation}
which facilitates the solution of the constraints.
According to the embedding of $SU(2) \times U(1)$ in $E_{8}$, its generators can be divided as
\begin{equation}\label{gens1}
Q_{E_{8}} = \{ G^{\rho}, G, G^{\alpha}, G^{a}, G_{a}, G^{i}, G_{i},
G^{ai}, G_{ai} \}
\end{equation}
where, $\rho = 1,2,3$, $a=1,2$, $\alpha=1,\ldots,78$,
$i=1,\ldots,27$.
Then we can write the solutions of the constraints (\ref{constraints}) in
terms of the genuine Higgs fields $\beta^{i}$, $\gamma^{i}$ and the
$E_{8}$ generators (\ref{gens1}) corresponding to the embedding
(\ref{dec4}) as follows,
\begin{eqnarray}
\phi^{\rho}&=&G^{\rho}, ~~~~~~~~~~~~~~~~~\phi=\sqrt{3}G, \nonumber \\
\phi_{a}&=&R_{1}\frac{1}{\sqrt{2}}\beta^{i}G_{1i},~~~~~
\phi_{+}=R_{2}\gamma^{i}G_{i}.
\end{eqnarray}

The four-dimensional potential in terms of the physical scalar fields
$\beta^{i}$ and $\gamma^{i}$ becomes \cite{Manousselis:2000aj}
\begin{eqnarray}\label{gpotex2}
V(\beta^{i},\gamma^{i})&=&\frac{\alpha'}{8\kappa^2}
e^{-\miso\t{\phi}}\biggl[\mbox{const}
-\frac{6}{R_{1}^{2}}\beta^{2}
-\frac{4}{R_{2}^{2}}\gamma^{2}+\bigg(4\sqrt{\frac{10}{7}}R_{2}
\bigl(\frac{1}{R_{2}^{2}}+\frac{1}{2R_{1}^{2}}\bigr)
d_{ijk}\beta^{i}\beta^{j}\gamma^{k} + h.c \bigg) \nonumber \\
&+&6\biggl(\beta^{i}(G^{\alpha})_{i}^{j}\beta_{j}
+\gamma^{i}(G^{\alpha})_{i}^{j}\gamma_{j}\biggr)^{2}
+\frac{1}{3}\biggl(\beta^{i}(1\delta_{i}^{j})\beta_{j}+
\gamma^{i}(-2\delta_{i}^{j})\gamma_{j}\biggr)^{2}\nonumber\\
&+&\frac{5}{7}\beta^{i}\beta^ {j}d_{ijk}d^{klm}\beta_{l}\beta_{m}
+4\frac{5}{7}\beta^{i}\gamma^{j}d_{ijk}d^{klm}\beta_{l}\gamma_{m}\biggl]vol_4.
\end{eqnarray}
In the last expression we have adopted our redefinitions (\ref{redef2n}) and the radii $R_1$ and $R_2$ are given by eq. (\ref{radii2}).

In the potential (\ref{gpotex2}) we observe again the appearance of the $F$- and $D$-terms and moreover the possible soft supersymmetry breaking terms.

\

\underline{Three-form~sector}

\

In order to determine the potential arising from the three-form sector in this example we write down the general expression (\ref{hpotentialgen}) keeping track of the number of scalar fields:
\bea  V_{H}=\frac{1}{4\kappa^2}
        e^{-\t{\phi}}\biggl[&&(2b^1+b^2)^2  - \frac{2\alpha'}{3}(\epsilon_1^{abc}b^1+\epsilon_2^{abc}b^2)Tr(\phi_a\phi_b\phi_c)\nn\\
        &&+\frac{\alpha'}{2}(\epsilon_1^{abc}b^1+\epsilon_2^{abc}b^2)Tr(f_{ab}^d\phi_c\phi_d)
            +\frac{2\alpha'^2}{3}Tr(\phi_{a}\phi_b\phi_{c})^2\nn\\&&+
            \frac{\alpha'^2}{16}Tr(f_{ab}^d\phi_{c}\phi_d)Tr(f_{[ab}^d\phi_{c]}\phi_d)
         -\alpha'^2Tr(\phi_a\phi_b\phi_c)Tr(f_{ab}^d\phi_c\phi_d)\biggl]vol_4,\nn\\
\end{eqnarray}
where we have substituted the value of the constants $n_1$ and $n_2$, which are given in Appendix B. As before we would like to express the potential in terms of the genuine
Higgs fields by using the same complex scalars we defined for the
gauge sector in eq. (\ref{redefex2}). The result that we obtain is
\begin{eqnarray}\label{hpotex2}
V_{H} = \frac{1}{4\kappa^2}
e^{-\t{\phi}}\biggl[&&\frac{1}{(R_1^2R_2)^2}(2{b^1}+{b^2})^2+\sqrt{2}i\alpha'\frac{1}{R_1^2R_2}(2{b^1}
        +{b^2})(d_{ijk}\beta^i\beta^j\gamma^k-h.c.)
            \nn\\&&+8\alpha'^2\beta^i\beta^j\gamma^kd_{ijk}d^{lmn}\beta_l\beta_m\gamma_n
        +\alpha'^2(\frac{\beta^2}{R_1}+\frac{\gamma^2}{R_2})^2\nn\\
            &&+\sqrt{6}\alpha'^2(\frac{\beta^2}{R_1}+\frac{\gamma^2}{R_2})(d_{ijk}\beta^i\beta^j\gamma^k+h.c.)\biggl]vol_4.
\eea

\

\subsection{Example based on ${\3}$}

\

{\it{Gravity sector and the nearly-K\"{a}hler limit}}

\

In the present case of $\3$ we perform the following redefinition of the scalar moduli fields $\phi,\vf,\chi$ and $\psi$, \bea\label{redef3n} \t{\phi} &=& -\miso(\phi+4\alpha\varphi),\nn \\
            \t{\varphi} &=& -\miso(\phi-\frac{4\alpha}{3}\varphi+4\gamma\chi+4\delta\psi), \nn\\
             \t{\chi} &=& -\miso(\phi-\frac{4\alpha}{3}\varphi+4\gamma\chi-4\delta\psi),
            \nn \\ \t{\psi} &=& -\miso(\phi-\frac{4\alpha}{3}\varphi-8\gamma\chi).
            \eea
Then the potential (\ref{potex3}) can be written in terms of the redefined scalar fields in the form \beq\label{potex3r} V_{grav} =-\frac{1}{4\kappa^2}{e^{-\t{\phi}}}(\frac{6}{R_1^2}+\frac{6}{R_2^2}+\frac{6}{R_3^2}-\frac{R_1^2}{R_2^2R_3^2}
        -\frac{R_2^2}{R_1^2R_3^2}-\frac{R_3^2}{R_1^2R_2^2}),\eeq where the radii $R_1(x),R_2(x)$ and $R_3(x)$ of $\3$ are given by \bea\label{radii3} R_1^2&=&R^2e^{-{\t{\vf}}}, \nn\\
        R_2^2&=&R^2e^{-\t{\chi}}, \nn\\
        R_3^2&=&R^2e^{-\t{\psi}}.
\eea In the same spirit as before, the nearly-K\"{a}hler limit in the case of $\3$, which amounts to setting the radii equal, $R_1=R_2=R_3$, is obtained when $\t{\vf}=\t{\chi}=\t{\psi}$. It is interesting to note that the limiting form of the potential (\ref{potex3r}) is formally given again by the expression (\ref{potlimit}).

\

\

{\it{Gauge sector}}

\

Concerning the gauge sector of this model, the only difference as compared to the previous ones
is that the chosen coset space to reduce the same theory is the $\3$. The decompositions to be
used are $$ E_{8} \supset SU(2) \times U(1) \times E_{6} \supset
U(1) \times U(1) \times E_{6} $$ The $248$ of $E_{8}$ is decomposed
under $SU(2) \times U(1)$ according to (\ref{dec4}), whereas the
decomposition under $U(1) \times U(1)$ is the following:
\begin{eqnarray}\label{fdec}
 248 &=& 1_{(0,0)}+1_{(0,0)}+1_{(1,3)}+1_{(-1,3)}\nonumber\\
&+&1_{(2,0)}+1_{(-2,0)}+1_{(-1,-3)}+1_{(1,-3)}\nonumber\\
&+&78_{(0,0)}+27_{(1,1)}+27_{(-1,1)}+27_{(0,-2)}\nonumber\\
&+&\overline{27}_{(-1,-1)}+\overline{27}_{(1,-1)}
+\overline{27}_{(0,2)}.
\end{eqnarray}
In the present case $R$ is chosen to be identified with the $U(1)
\times U(1)$ of the latter decomposition. Therefore the resulting
four-dimensional gauge group is $$ H=C_{E_{8}}(U(1) \times U(1)) =
 U(1) \times U(1) \times E_{6}.$$ Here applies the same remark as in the previous case, namely the extra $U(1)$s in the four-dimensional gauge group do not correspond to any local symmetry. Hence we focus again on the $E_6$ part of the gauge group.  The $R=U(1) \times U(1)$
content of the $\3$ vector and spinor, according
to tables 5 and 6, are $$(1,1)+(-1,1) +(0,-2)+(-1,-1)+(1,-1)+(0,2)$$ and
$$(0,0)+(1,1)+(-1,1) +(0,-1)$$ respectively. Thus applying the CSDR rules one finds that
the surviving fields in four dimensions are one ${\cal N}=1$
vector multiplet $V^{\alpha}$, where $\alpha$ is
an $E_{6}$ $78$ index, and three ${\cal N}=1$ chiral multiplets
($A^{i}$, $B^{i}$, $\Gamma^{i}$) with $i$ an $E_{6}$ $27$ index. The number of fermion families in the 27 of $E_6$ is three, as expected since the coset space has Euler characteristic six.

To determine the potential the decomposition of
the adjoint of the specific $S=SU(3)$ under $R=U(1) \times U(1)$ has to be examined,
i.e.
$$SU(3) \supset U(1) \times U(1) $$
\begin{eqnarray}\label{decsu}
 8 = (0,0)+(0,0)+(1,1)+(-1,1)
+(0,-2)+\nonumber\\(-1,-1)+(1,-1)+(0,2).
\end{eqnarray}
Then according to the decomposition (\ref{decsu}) the generators of
$SU(3)$ can be grouped as
\begin{equation}\label{sugen}
Q_{SU(3)} = \{Q_{0},Q'_{0},Q_{1},Q_{2},Q_{3},Q^{1},Q^{2},Q^{3} \}.
\end{equation}
The decomposition
(\ref{decsu}) suggests the following change in  the notation of the
scalar fields,
\begin{equation}\label{redefex3}
(\phi_{I}, I=1,\ldots,8) \longrightarrow ( \phi_{0}, \phi'_{0},
\phi_{1}, \phi^{1}, \phi_{2}, \phi^{2}, \phi_{3}, \phi^{3}).
\end{equation}
Moreover, under the decomposition (\ref{fdec}) the generators
of $E_{8}$ can be grouped as \beq\label{fgens}
Q_{E_{8}}=\{Q_{0},Q'_{0},Q_{1},Q_{2},Q_{3},Q^{1},Q^{2},Q^{3},Q^{\alpha},
Q_{1i},Q_{2i},Q_{3i},Q^{1i},Q^{2i},Q^{3i} \}, \eeq where, $
\alpha=1,\ldots,78 $ and $ i=1,\ldots,27 $.
Then the constraints (\ref{constraints}) are solved according to
\begin{eqnarray}\label{fcons}
\phi_{1} &=& R_{1} \alpha^{i} Q_{1i}, \nonumber\\
\phi_{2} &=& R_{2} \beta^{i} Q_{2i}, \nonumber\\
\phi_{3} &=& R_{3} \gamma^{i} Q_{3i},
\end{eqnarray}
where the unconstrained  scalar fields transform under $27$ of $E_{6}$.
Then the potential is
expressed in terms of the genuine Higgs fields as
\begin{eqnarray}\label{gpotex3}
V(\alpha^{i},\beta^{i},\gamma^{i})&=& \frac{\alpha'}{8\kappa^2}e^{-\miso\t{\phi}}\biggr[\mbox{const.} + \biggl(
\frac{4R_{1}^{2}}{R_{2}^{2}R_{3}^{2}}-\frac{8}{R_{1}^{2}}
\biggr)\alpha^{i}\alpha_{i}
+\biggl(\frac{4R_{2}^{2}}{R_{1}^{2}R_{3}^{2}}-\frac{8}{R_{2}^{2}}\biggr)
\beta^{i}\beta_{i} \nn\\&+&\biggl(\frac{4R_{3}^{2}}{R_{1}^{2}R_{2}^{2}}
-\frac{8}{R_{3}^{2}}\biggr)\gamma^{i}\gamma_{i}
+\sqrt{2}80\biggl(\frac{R_{1}}{R_{2}R_{3}}+\frac{R_{2}}{R_{1}
R_{3}}+\frac{R_{3}}{R_{2}R_{1}}\biggr)(d_{ijk}\alpha^{i}\beta^{j}\gamma^{k}+h.c.)
\nn\\&+&\frac{1}{6}\biggl(\alpha^{i}(G^{\alpha})_{i}^{j}\alpha_{j}
+\beta^{i}(G^{\alpha})_{i}^{j}\beta_{j}
+\gamma^{i}(G^{\alpha})_{i}^{j}\gamma_{j}\biggr)^{2}\nonumber\\
&+&\frac{10}{6}\biggl(\alpha^{i}(3\delta_{i}^{j})\alpha_{j}  +
\beta^{i}(-3\delta_{i}^{j})\beta_{j}
 \biggr)^{2}
+\frac{40}{6}\biggl(\alpha^{i}(\frac{1}{2}\delta_{i}^{j})\alpha_{j}
+ \beta^{i}(\frac{1}{2}\delta^{j}_{i})\beta_{j} +
\gamma^{i}(-1\delta_{i}^{j})\gamma_{j}
 \biggr)^{2}\nonumber \\
&+&40\alpha^{i}\beta^{j}d_{ijk}d^{klm}\alpha_{l}\beta_{m}
+40\beta^{i}\gamma^{j}d_{ijk}d^{klm}\beta_{l}\gamma_{m}
+40\alpha^{i}\gamma^{j}d_{ijk}d^{klm}\alpha_{l}\gamma_{m}\biggr]vol_4,\nonumber\\
\end{eqnarray}
where $R_{1},R_{2},R_{3}$ are the coset space radii as defined in (\ref{radii3}).

\

{\it{Three-form~sector}}

\

In order to determine the potential arising from the three-form sector in this example we write down again the general expression (\ref{hpotentialgen}) keeping track of the number of scalar fields:
\bea  V_{H}=\frac{1}{4\kappa^2}
        e^{-\t{\phi}}\biggl[&&(b^1+b^2+b^3)^2  - \frac{2\alpha'}{3}(\epsilon_1^{abc}b^1+\epsilon_2^{abc}b^2+\epsilon_3^{abc}b^3)Tr(\phi_a\phi_b\phi_c)\nn\\
        &&+\frac{\alpha'}{2}(\epsilon_1^{abc}b^1+\epsilon_2^{abc}b^2+\epsilon_3^{abc}b^3)Tr(f_{ab}^d\phi_c\phi_d)
            +\frac{2\alpha'^2}{3}Tr(\phi_{a}\phi_b\phi_{c})^2\nn\\&&+
            \frac{\alpha'^2}{16}Tr(f_{ab}^d\phi_{c}\phi_d)Tr(f_{[ab}^d\phi_{c]}\phi_d)
            -\alpha'^2Tr(\phi_a\phi_b\phi_c)Tr(f_{ab}^d\phi_c\phi_d)\biggl]vol_4,\nn\\
\end{eqnarray}
where we have substituted the value of the constants $n_1$ and $n_2$, which are given in Appendix B. Expressing the potential in terms of the genuine
Higgs fields by using the same complex scalars we defined for the
gauge sector in eq. (\ref{redefex3}), we obtain the result
\begin{eqnarray}\label{hpotex3}
V_{H} = \frac{1}{4\kappa^2}
e^{-\t{\phi}}\biggl[&&\frac{1}{(R_1R_2R_3)^2}({b^1}+{b^2}+{b^3})^2
    +\sqrt{2}i\alpha'\frac{1}{R_1R_2R_3}({b^1}+{b^2}+{b^3})(d_{ijk}\alpha^i\beta^j\gamma^k-h.c.)
            \nn\\&&+8\alpha'^2\alpha^i\beta^j\gamma^kd_{ijk}d^{lmn}\alpha_l\beta_m\gamma_n
        +\alpha'^2(\frac{\alpha^2}{R_1}+\frac{\beta^2}{R_2}+\frac{\gamma^2}{R_3})^2\nn\\
            &&+\sqrt{6}\alpha'^2(\frac{\alpha^2}{R_1}+\frac{\beta^2}{R_2}+\frac{\gamma^2}{R_3})(d_{ijk}\alpha^i\beta^j\gamma^k+h.c.)\biggl]vol_4.
\eea

\section{Supergravity description in four dimensions}

\subsection{Generalities}

Having determined the four-dimensional theory in the previous sections we would like to attempt to provide a supergravity description in four dimensions. The bosonic sector of the ${\cal N}=1$, four-dimensional supergravity is given in terms of the Lagrangian \cite{Wess:1992cp},
\beq\label{sugra4} {\cal
L}_b=-\frac{1}{2\kappa^2}R*1-\frac{1}{2}Re(f)F^I\w\ast
F^I+\frac{1}{2}Im(f)F^I\w {F}^I-\frac{1}{\kappa^2}G_{i\bar{j}}d\Phi^i\w\ast
d\bar{\Phi}^{\bar{j}}-V(\Phi,\overline{\Phi}), \eeq where $G_{i\bar{j}}$ is the K\"{a}hler
metric, determined by the K\"{a}hler potential through the formula,
\beq
G_{i\bar{j}}=\frac{\partial}{\partial\Phi^i}\frac{\partial}{\partial\bar{\Phi}^{\bar{j}}}K(\Phi,\bar{\Phi})
\eeq and by $\Phi^i$ we collectively denote the chiral multiplets.
Moreover, the potential has the form
\beq\label{spot}
V(\Phi,\bar{\Phi})=\frac{1}{\kappa^4}e^{\kappa^2K}\big(K^{i\bar{j}}
\frac{DW}{D\Phi^i}\frac{D\overline{W}}{D\bar{\Phi}^{\bar{j}}}-3\kappa^2W\overline{W}\big)+D-\mbox{terms},
\eeq where the derivatives involved are the K\"{a}hler covariant
derivatives
 \beq \frac{DW}{D\Phi^i}=\frac{\partial
W}{\partial\Phi^i}+\frac{\partial K}{\partial\Phi^i}W.\eeq

Thus, in order to express the reduced Lagrangian we determined in the
standard ${\cal N}=1$ form in four dimensions we have to specify the
gauge kinetic function, $f$, the K\"{a}hler potential, K, and the
superpotential, W.

In order to determine the superpotential of the four-dimensional theory we shall employ the Gukov-Vafa-Witten formula \cite{Gukov:1999ya},\cite{Gukov:1999gr}, which has the form
\beq\label{gukov} W=\frac{1}{4}\int_{S/R}\Omega\w(\hat{H}+idJ), \eeq
and it was shown to be the appropriate formula for general heterotic compactifications on manifolds with $SU(3)$-structure in \cite{Gurrieri:2004dt}, \cite{Benmachiche:2008ma}, \cite{LopesCardoso:2003af}.

The K\"{a}hler potential can be determined as the sum of two terms{\footnote{A third term which is in general associated to the complex structure moduli is absent in our formalism, since the SU(3) structures we are considering
on these manifolds are not complex.}}, \beq\label{kahlerpot} K=K_S+K_{T}, \eeq which are given by the expressions \bea K_S &=& -ln(S+{S}^*),\\
                    K_T &=& -ln{\cal K}, \eea where $S$ is the superfield involving the scalars $\t{\phi}$ and $\theta$ in the combination \beq S= e^{\t{\phi}}+i\theta \eeq and ${\cal K}$ is the volume of the internal manifold, given by the expression \beq {\cal K} = \frac{1}{6}\int_{S/R}J\w J\w J. \eeq Let us note that comparing eqs. (\ref{gaugel}), (\ref{RL4}) and (\ref{sugra4}) we can immediately conclude that the gauge kinetic function is $f(S)=S$ in all cases.

\subsection{The ${\1}$ case}

There exist four scalar moduli fields in four dimensions resulting from the dimensional reduction in this case, namely $\t{\phi},\t{\varphi},\theta$ and $b$. There exist also the additional scalar
fields $\beta^i$ arising from the internal components of the higher-dimensional gauge field.

In order to determine the four-dimensional superpotential from eq. (\ref{gukov}) we have to use the general expression for $\hat{H}$ in eq. (\ref{hhat}) and the expressions for $\Omega$ and $dJ$ which can be found in appendix B. A direct calculation leads to the result
\beq W={3}T_1-\sqrt{2}\alpha'd_{ijk}B^iB^jB^k, \eeq
where we have defined the superfields $T_1$ and $B^i$. The $T_1$ involves the scalar fields $\t{\vf},b$ and $\beta$ in the combination \beq T_1=e^{{-\frac{\t{\varphi}}{\sqrt{3}}}}+ib+\alpha'\beta^2, \eeq while by $B^i$ we denote the superfields whose scalar components are the fields $\beta^i$.

Moreover, we determine the K\"{a}hler potential using (\ref{kahlerpot}) and we find that it takes the form
\beq
K=-ln(S+S^*)-3ln(T_1+T_1^*-2\alpha'B_iB^i). \eeq

With these K\"{a}hler potential and superpotential eq. (\ref{spot}) reproduces the four-dimensional potential we have determined through dimensional reduction, namely the three contributions from the gravity, gauge and three-form sectors appearing in eqs. (\ref{potex1r}), (\ref{pot44}) and (\ref{hpotex1}) respectively. In addition, all the kinetic terms we have determined are exactly retrieved as \beq
-\frac{1}{\kappa^2}G_{i\bar{j}}d\Phi^i\w\ast d\bar{\Phi}^{\bar{j}}
\eeq with the same K\"{a}hler potential, as required by
supergravity.

\subsection{The ${\2}$ case}

There exist six scalar moduli fields in this case, namely $\t{\phi},\theta,\t{\varphi},\t{\chi},b^1$ and $b^2$. In addition two more sets of scalar
fields $\beta^i$ and $\gamma^i$ arise from the internal components of the higher-dimensional gauge field.

Calculating the superpotential from eq. (\ref{gukov}) we obtain the result \beq W=2T_1+T_2-\sqrt{2}\alpha'd_{ijk}B^iB^j\Gamma^k, \eeq where we have defined the superfields
\bea
T_1&=&e^{-\t{\varphi}/\sqrt{2}}+ib^1+\alpha'\beta^2,\nn\\T_2&=&e^{-\t{\chi}}+ib^2+\alpha'\gamma^2, \eea while by $B^i$ we denote the superfields whose scalar components are the fields $\beta^i$ and by $\Gamma^i$ the corresponding ones with scalar components $\gamma^i$.

The K\"{a}hler potential can be determined again using the expression (\ref{kahlerpot}) and it takes the form,
\beq
K=-ln(S+S^*)-2ln(T_1+T_1^*-2\alpha'B_iB^i)-ln(T_2+T_2^*-2\alpha'\Gamma_i\Gamma^i) \eeq

With the above K\"{a}hler potential and superpotential eq. (\ref{spot}) reproduces again the four-dimensional potential we have determined through dimensional reduction, namely the contributions appearing in eqs. (\ref{potex2r}), (\ref{gpotex2}) and (\ref{hpotex2}). In addition, all the kinetic terms we have determined are again exactly reproduced as in the previous case. Finally let us note that the nearly-K\"{a}hler limit is obtained when $T_1=T_2$.

\subsection{The ${\3}$ case}

Here the number of scalar moduli is eight, namely $\t{\phi},\theta,\t{\varphi},\t{\chi},\t{\psi},b^1,b^2$ and $b^3$. There exist also now three sets of additional scalar
fields $\alpha^i,$$\beta^i$ and $\gamma^i$ arising from the internal components of the higher-dimensional gauge field.
Eq. (\ref{gukov}) leads in the present case to the superpotential
\beq W=T_1+T_2+T_3-\sqrt{2}\alpha'd_{ijk}A^iB^j\Gamma^k, \eeq where the superfields appearing in this expression are now defined as
\bea
T_1&=&e^{-\t{\varphi}}+ib^1+\alpha'\alpha^2,\\T_2&=&e^{-\t{\chi}}+ib^2+\alpha'\beta^2,
\\T_3&=&e^{-\t{\psi}}+ib^3+\alpha'\gamma^2\eea
and by $A^i,B^i$ and $\Gamma^i$ we denote again the superfields whose scalar components are the corresponding scalar fields.

Eq. (\ref{kahlerpot}) yields for the K\"{a}hler potential the result
\beq
K=-ln(S+S^*)-ln(T_1+T_1^*-2\alpha'A_iA^i)-ln(T_2+T_2^*-2\alpha'B_iB^i)-ln(T_3+T_3^*-2\alpha'\Gamma_i\Gamma^i). \eeq

With these K\"{a}hler potential and superpotential eq. (\ref{spot}) reproduces again the four-dimensional potential we have determined through dimensional reduction, namely the three contributions appearing in eqs. (\ref{potex3r}), (\ref{gpotex3}) and (\ref{hpotex3}), as well as all the kinetic terms appearing in the four-dimensional theory. Finally, the nearly-K\"{a}hler limit corresponds to $T_1=T_2=T_3$.

\section{Conclusions}

In the present work we have explicitly reduced the heterotic
supergravity coupled to super Yang-Mills from ten dimensions to four at first order in
$\alpha'$ using homogeneous six-dimensional nearly-K\"{a}hler manifolds as
internal spaces. Since the homogeneous nearly-K\"{a}hler manifolds in six dimensions are the three corresponding non-symmetric coset spaces plus a group manifold, we employed the Coset Space Dimensional Reduction scheme to reduce the gauge sector of the theory. In our discussion we excluded the group manifold case since it does not meet the requirement of obtaining chiral fermions in four dimensions. Concerning the reduction of the other parts of the ten-dimensional theory we provided appropriate ansatze which amount to the expansion of the fields involved in $S$-invariant $p$-forms of the internal manifolds. Subsequently, we determined the general form of the four-dimensional Lagrangian obtained by the dimensional reduction of the bosonic Lagrangian of the theory over the non-symmetric coset spaces. Next we determined the full four-dimensional potential in terms of the surviving scalar fields for the three homogeneous nearly-K\"{a}hler manifolds, namely $\1,\2$ and $\3$. This potential contains terms which could be interpreted as soft scalar masses and trilinear soft terms in four-dimensions, in case the minimization of the full potential would lead to Minkowski vacuum. This possibility hopefully is not excluded if all possible condensates are taken into account, while some uplifting mechanisms have been already proposed.

Finally, attempting a supergravity description of our results from the four-dimensional viewpoint we have employed the Gukov-Vafa-Witten formula for the superpotential as well as the formulae for the K\"{a}hler potential which are appropriate when the internal space is an $SU(3)$-structure manifold. Using the forementioned formulae we have determined the superpotential and K\"{a}hler potential in all cases, which can reproduce the four-dimensional potential and kinetic terms.

\

\noindent {\bf Acknowledgments}

\

We would like to thank P. Aschieri, G.L. Cardoso, A. Kehagias, C.
Kounnas, G. Koutsoumbas, J. Louis, D. L\"{u}st, P. Manousselis and D. Tsimpis for useful
discussions. This work is supported by the NTUA programme for basic research 'PEVE 2008'
and the European Union's RTN programme under contract
MRTN-CT-2006-035505.
\vspace{2cm}

\noindent {\bf\Large Appendix}

\appendix

\section{Geometry of coset spaces}

The geometry of coset spaces $S/R$ relevant for our purposes is
presented in refs.~\cite{Castellani:1983tb, Castellani:1999fz}. Let
the coordinates of the Lie group $S$ be $(y^{a}, z^{i})$ with
$y^{a}$ being the coset coordinates and $z^{i}$ being the
coordinates of the $R$ subgroup. Then a group element $s \in S$ can
be represented as $s \sim e^{y^{a}Q_{a}} e^{z^{i}Q_{i}}$ and a coset
representative is $L(y) = e^{y^{a}Q_{a}}$. The Maurer--Cartan 1-form
is defined by $e(y)=L^{-1}(y)dL$ and is the analogue of the
left-invariant 1-form on a Lie group $S$. It takes values in
$Lie(S)$, i.e.~the Lie algebra of $S$:
\begin{equation}
e(y) = e^{A}Q_{A}= e^{a}Q_{a} + e^{i}Q_{i},
\end{equation}
where $A$ is a group index, $e^{a}$ is the coframe and $e^{i}$ is
the $R$-connection. The latter can be expanded in coset vielbeins as
$e^i=e^i_a(y) e^a$. The exterior derivative of the Maurer--Cartan
1-form is
\begin{equation}
de = d(L^{-1}dL) = - e \wedge e = - [e,e],
\end{equation}
from which we can easily prove that
\begin{equation}
de^{A} = - \frac{1}{2}f^{A}_{BC} e^{B} \wedge e^{C}.\label{mcg}
\end{equation}

We will assume, for reasons analyzed in detail in
ref.~\cite{Castellani:1983tb}, that the coset is reductive. That
means that the commutation relations obeyed by the generators of $S$
are not the most general ones but they take the form
\begin{eqnarray}
\left[ Q_{i}, Q_{j} \right] &=& f_{ij}^{k} Q_{k}, \nonumber \\
\left[Q_{i}, Q_{a}\right] &=& f_{ia}^{b}Q_{b}, \nonumber \\
\left[Q_{a},Q_{b}\right]&=& f_{ab}^{c} Q_{c} + f_{ab}^{i} Q_{i},
\end{eqnarray}
implying that  $f_{bi}^{j}=0$. Now (\ref{mcg}) can be written as
\begin{eqnarray}
de^{a} = - \frac{1}{2}f^{a}_{ bc}e^{b} \wedge e^{c} -
f^{a}_{ bi}e^{b} \wedge e^{i}, \label{first} \\
de^{i} = -\frac{1}{2} f^{i}_{ab}e^{a} \wedge e^{b} -
\frac{1}{2}f^{i}_{ jk}e^{j} \wedge e^{k}
\end{eqnarray}
and from eq.~(\ref{first}) we can obtain the Maurer--Cartan
equations for the coset vielbeins
\begin{equation}\label{second}
de^{a} = - \frac{1}{2}C^{a}_{ bc}(y) e^{b} \wedge e^{c}, \;\;\;
C^{a}_{ bc} = f^{a}_{ bc} - 2e^{i}_{[b}f^{a}_{c]i}.
\end{equation}

Finally, an $S$-invariant metric on $S/R$ is
\begin{equation}\label{Smetric}
g_{\alpha\beta}(y) = \delta_{a b}e^{a}_{\alpha}(y)e^{b}_{\beta}(y).
\end{equation}
Using the metric (\ref{Smetric}) the following useful identities can
be proved
\begin{eqnarray}\label{ident}
&&e^{a} \wedge \ast_{d} e^{b} = \delta^{ab} vol_{d},\\
&&(e^a \wedge e^b) \wedge *_{d} (e^c \wedge e^d) = \delta^{ab}_{cd}
vol_{d},\\
&&(e^a \wedge e^b \wedge e^c) \wedge *_{d} (e^d \wedge e^e \wedge
e^f) = \delta^{abc}_{def} vol_{d}.
\end{eqnarray}
where $\ast_{d}$ is the Hodge duality operator on a $d$-dimensional
coset.

\section{Data for the coset spaces}

In this appendix we provide all the data related to the internal manifolds we use in the process of reduction of the
heterotic supergravity coupled to super Yang-Mills from ten to four dimensions. Specifically, we
collect the metric, structure constants and invariant forms of these
spaces, as well as some characteristic constants of them which appear
in the main text. Detailed tables of the field content of the theory obtained in each case are also given.

\subsection*{$\1$}

$\bullet$ Metric:
\begin{equation}
ds^2=R_1^2 (e^1 \otimes e^1 + e^2 \otimes e^2 + e^3 \otimes e^3 + e^4
\otimes e^4 + e^5 \otimes e^5 + e^6 \otimes e^6).
\end{equation}
$\bullet$ Structure Constants (7-14 correspond to the generators of
$SU(3)$): \bea
f_{136}&=&f_{145}=-f_{235}=f_{246}=\frac{1}{\sqrt{3}},\nn\\
2f_{789}&=&f_{7,10,13}=-f_{7,11,12}=f_{736}=-f_{745}=f_{8,10,12}=f_{8,11,13}=-f_{835}=-f_{846}=f_{9,10,11}\nn\\
&=&-f_{9,12,13}=-f_{934}=f_{956}=f_{10,1,6}=f_{10,2,5}=-f_{11,1,5}=f_{11,2,6}=f_{12,1,4}=f_{12,2,3}\nn\\
&=&-f_{13,1,3}=f_{13,2,4}=\miso,\nn\\
f_{10,11,14}&=&f_{12,13,14}=\frac{3}{2}f_{14,1,2}=3f_{14,3,4}=3f_{14,5,6}=\frac{\sqrt{3}}{2}.
\eea $\bullet$ Euler characteristic: $\chi = 2$.\\
$\bullet$ Invariant forms: \bea 2-\mbox{form}: \omega_1 &=& e^{12}-e^{34}-e^{56}. \\
                                3-\mbox{forms}: \rho_1 &=&
                                            e^{136}+e^{145}-e^{235}+e^{246},
                                           ~~~ \rho_2 =
                                            e^{135}-e^{146}+e^{236}+e^{245}.
                                            \eea
$\bullet$ SU(3)-structure: \bea J &=& R_1^2\omega_1,\\
                                dJ &=& -\sqrt{3}R_1^2\rho_2,\\
                                \Omega &=& R_1^3(\rho_2+i\rho_1),\\
                                d\Omega &=&
                                \frac{8i}{\sqrt{3}}R_1^3(e^{1234}+e^{1256}-e^{3456}).
                                \eea
$\bullet$ Intrinsic torsion class: \beq {\cal
W}_1=-\frac{2i}{\sqrt{3}R_1}. \eeq $\bullet$ Data related to
(\ref{data}):
\bea m&=& 3, \nn\\  n_1&=& 12, \nn\\
 \epsilon_1^{ab}&=&\delta^{ab}_{12}-\delta^{ab}_{34}-\delta^{ab}_{56},\nn\\
  \epsilon_1^{abc}&=&-\sqrt{3}(\delta^{abc}_{135}-\delta^{abc}_{146}+\delta^{abc}_{236}+\delta^{abc}_{245}).  \eea
\newpage
$\bullet$ Tables of field decompositions:

\

\begin{center}
\begin{tabular}{|c|c|c|}\hline \multicolumn{3}{|c|}{Table 1}\\
\multicolumn{3}{|c|}{Decomposition of bosonic fields under $SU(3)$}
\\ \hline
\hline field & components & $SU(3)~\mbox{representations}$
\\
\hline \multirow{3}{*}{$\hat G_{MN}$} & $g_{\mu\nu}$ & $\mathbf{1}$
\\
\cline{2-3} & $g_{\mu m}$ & $\mathbf{3}+\mathbf{\bar 3}$
\\
\cline{2-3}
& $g_{mn}$ & $\mathbf{1} + \mathbf{6} + \mathbf{\bar 6} + \mathbf{8}$ \\
\hline \multirow{3}{*}{$\hat B_{MN}$} & $B_{\mu\nu}$ & $\mathbf{1}$
\\ \cline{2-3} & $B_{\mu m}$ &
$\mathbf{3}+\mathbf{\bar 3}$
\\ \cline{2-3}
& $B_{mn}$ & $\mathbf{1} + \mathbf{3} + \mathbf{\bar 3} + \mathbf{8}$ \\
\hline
$\hat\phi$ & $\phi$ & $\mathbf{1}$ \\
\hline \multirow{2}{*}{$\hat A^{\hat A}_M$} & $A_\mu$ & $\mathbf{1}$
\\ \cline{2-3}
& $A_m$ & $\mathbf{3}+\mathbf{\bar 3}$ \\
\hline
\end{tabular}
\end{center}

\

\begin{center}
\begin{tabular}{|c|c|c|}\hline \multicolumn{3}{|c|}{Table 2}\\
\multicolumn{3}{|c|}{Decomposition of fermionic fields under
$SU(3)$} \\\hline field & components & $SU(3)~\mbox{representations}$ \\ \hline
\multirow{2}{*}{$\hat\psi_M$} & $\psi_\mu$ & $\mathbf{1} +
\mathbf{3}$ \\ \cline{2-3}
& $\psi_m$ & $\mathbf{1} + 2\cdot\mathbf{3} +\mathbf{\bar 3} + \mathbf{\bar 6} + \mathbf{8}$  \\
\hline
$\hat \lambda$ & $\lambda$ & $\mathbf{1} + \mathbf{3}$ \\
\hline
$\hat \chi$ & $\chi$ & $\mathbf{1} + \mathbf{3}$ \\
\hline
\end{tabular}
\end{center}

\

\subsection*{ ${\2}$}

$\bullet$ Metric:
\begin{equation}
ds^2=R_1^2 (e^1 \otimes e^1 + e^2 \otimes e^2) + R_2^2 (e^3 \otimes e^3 +
e^4 \otimes e^4) + R_1^2 (e^5 \otimes e^5 + e^6 \otimes e^6).
\end{equation}
$\bullet$ Structure Constants (7-10 correspond to the generators of
$SU(2)\times U(1)$): \bea
f_{136}&=&-f_{145}=f_{235}=f_{246}=\miso,\nn\\
2f_{789}&=&f_{716}=-f_{725}=f_{815}=f_{826}=f_{912}=-f_{956}=f_{10,1,2}=2f_{10,3,4}=f_{10,5,6}=\miso.\nn\\
\eea $\bullet$ Euler characteristic: $\chi = 4$.\\$\bullet$ Invariant forms: \bea \mbox{2-forms}: \omega_1 &=&
-e^{12}-e^{56},
                                            ~~~\omega_2 = e^{34}, \\
                                \mbox{3-forms}: \rho_1 &=&
                                            e^{136}-e^{145}+e^{235}+e^{246},
                                           ~~~ \rho_2 =
                                            e^{135}+e^{146}-e^{236}+e^{245}.
                                            \eea
$\bullet$ $SU(3)$-structure:
\begin{eqnarray}
J &=& R_1^2\omega_1+R_2^2\omega_2,\\
dJ&=& -(2 R_1^2 + R_2^2) \rho_2,\\
\Omega &=&{R_1^{2} R_2 }(\rho_2+i\rho_1),\\
d\Omega &=& 4i {R_1^2R_2}(e^{1234} - e^{1256} + e^{3456} ).
\end{eqnarray}
$\bullet$ Intrinsic torsion classes:
\begin{eqnarray}
{\cal W}_1 &=& -\frac{2i}{3} \frac{2R_1^2+R_2^2}{{R_1^2 R_2}}, \\
{\cal W}_2 &=& -\frac{4i}{3} \frac{1}{{R_1^2 R_2}}\left[ R_1^2(R_1^2 -R_2^2)
e^{12} - 2R_2^2(R_2^2 - R_1^2)e^{34} + R_1^2(R_1^2-R_2^2)e^{56} \right].
\end{eqnarray}
$\bullet$ Data related to (\ref{data}): \bea m&=&
\left\{\begin{array}{llc} 2, & i=j = 1,
\\  1, & i =j= 2,  \end{array}\right. \nn\\  n_1&=& \left\{\begin{array}{llc} 16, & i = 1,
\\  4, & i = 2,  \end{array}\right. \nn\\
n_2&=&8,\nn\\
 \epsilon_1^{ab}&=&\delta^{ab}_{12}+\delta^{ab}_{56},\nn\\
 \epsilon_2^{ab}&=&\delta^{ab}_{34},\nn\\
  \epsilon_1^{abc}&=&-2\epsilon_2^{abc}=\delta^{abc}_{135}+\delta^{abc}_{146}-\delta^{abc}_{236}+\delta^{abc}_{245}.  \eea

$\bullet$ Tables of field decompositions (the subscripts denote the $U(1)$ charge):

\

\begin{center}
\begin{tabular}{|c|c|c|}\hline \multicolumn{3}{|c|}{Table 3}\\
\multicolumn{3}{|c|}{Decomposition of bosonic fields under
$SU(2)\times U(1)$}
\\ \hline field & components & $SU(2)\times U(1)~\mbox{representations}$
\\
\hline \multirow{3}{*}{$\hat G_{MN}$} & $g_{\mu\nu}$ &
$\mathbf{1}_{\m{0}}$
\\
\cline{2-3} & $g_{\mu m}$ & $\m{1}_{\m{2}}+\m{1}_{\m{-2}}+\m{2}_{\m{1}}+\m{2}_{\m{-1}}$
\\
\cline{2-3}
& {$g_{mn}$} & $2\cdot\mathbf{1}_{\mathbf{0}}+ \mathbf{1}_{\m{4}}+ \mathbf{1}_{\m{-4}}+ \mathbf{2}_{\m{1} }+ \mathbf{2}_{\m{-1} }
+\m{2}_{\m{3}}+ \mathbf{2}_{\m{-3}}+\m{3}_{\m{0}}+ \mathbf{3}_{\m{2} }+ \mathbf{3}_{\m{-2} }$ \\
\hline \multirow{3}{*}{$\hat B_{MN}$} & $B_{\mu\nu}$ &
$\mathbf{1}_\mathbf{0}$ \\ \cline{2-3} & $B_{\mu m}$ &
$\m{1}_{\m{2}}+\m{1}_{\m{-2}}+\m{2}_{\m{1}}+\m{2}_{\m{-1}}$
\\ \cline{2-3}
& $B_{mn}$ & $2\cdot\mathbf{1}_{\mathbf{0}}+\mathbf{1}_{\m{2}}+\mathbf{1}_{\m{-2}}+\m{2}_{\m{1}}
+\mathbf{2}_{\m{-1}}+\mathbf{2}_{\m{3}}+\mathbf{2}_{\m{-3}}+ \mathbf{3}_\mathbf{0}$ \\
\hline
$\hat\phi$ & $\phi$ & $\mathbf{1}_\mathbf{0}$ \\
\hline \multirow{2}{*}{$\hat A^{\hat A}_M$} & $A_\mu$ &
$\mathbf{1}_\mathbf{0}$ \\ \cline{2-3}
& $A_m$ & $\m{1}_{\m{2}}+\m{1}_{\m{-2}}+\m{2}_{\m{1}}+\m{2}_{\m{-1}}$ \\
\hline
\end{tabular}
\end{center}

\

\begin{center}
\begin{tabular}{|c|c|c|}\hline \multicolumn{3}{|c|}{Table 4}\\
\multicolumn{3}{|c|}{Decomposition of fermionic fields under
$SU(2)\times U(1)$}
\\ \hline field & components &
$SU(2)\times U(1)~\mbox{representations}$ \\ \hline \multirow{2}{*}{$\hat\psi_M$} &
$\psi_\mu$ &
$\mathbf{1}_\mathbf{0}+\m{1}_\m{-2} + \mathbf{2}_\mathbf{1}$ \\
\cline{2-3}
& $\psi_m$ & $2\cdot\m{1}_{\m{0}} +\mathbf{1}_{\m{2}}+\mathbf{1}_{\m{-2}}+\mathbf{1}_{\m{4}}+3\cdot\m{2}_{\m{1}}+\mathbf{2}_{\m{-1}}
+\mathbf{2}_{\m{3}}+\mathbf{2}_{\m{-3}}+\m{3}_{\m{0}}+\m{3}_\m{{-2}}$ \\
\hline
$\hat \lambda$ & $\lambda$ & $\mathbf{1}_\mathbf{0}+\m{1}_\m{-2} + \mathbf{2}_\mathbf{1}$ \\
\hline
$\hat \chi$ & $\chi$ & $\mathbf{1}_\mathbf{0}+\m{1}_\m{-2} + \mathbf{2}_\mathbf{1}$ \\
\hline
\end{tabular}
\end{center}


\

\subsection*{${\3}$}

$\bullet$ Metric:
\begin{equation}
ds^2=R_1^2 (e^1 \otimes e^1 + e^2 \otimes e^2) + R_2^2 (e^3 \otimes e^3 +
e^4 \otimes e^4) + R_3^2 (e^5 \otimes e^5 + e^6 \otimes e^6).
\end{equation}
$\bullet$ Structure Constants (3 and 8 correspond to the $U(1)\times
U(1)$ generators): \bea
2f_{123}&=&f_{147}=-f_{156}=f_{246}=f_{257}=f_{345}=-f_{367}=\miso,\nn\\f_{458}&=&f_{678}=\miso\sqrt{3}.
\eea $\bullet$ Euler characteristic: $\chi = 6$.\\$\bullet$ Invariant forms: \bea \mbox{2-forms}: \omega_1 &=& -e^{12},
                                            ~~~\omega_2
                                            = e^{45}, ~~~\omega_3 =
                                            -e^{67}. \\
                                \mbox{3-forms}: \rho_1 &=&
                                            e^{147}-e^{156}+e^{246}+e^{257},
                                           ~~~ \rho_2 =
                                            e^{146}+e^{157}-e^{247}+e^{256}.
                                            \eea
$\bullet$ $SU(3)$-structure:
\begin{eqnarray}
J &=& R_1^2 \omega_1+ R_2^2 \omega_2+ R_3^2\omega_3, \\
dJ &=& -(R_1^2 + R_2^2 + R_3^2) \rho_2, \\
\Omega &=& {R_1R_2R_3}(\rho_2+i\rho_1),\\
d\Omega &=& 4i {R_1R_2R_3} ( e^{1234}  - e^{1256}  +  e^{3456} ).
\end{eqnarray}
$\bullet$ Intrinsic torsion classes:
\begin{eqnarray}
{\cal W}_1 &=& -\frac{2i}{3} \frac{R_1^2+R_2^2+R_3^2}{{R_1R_2R_3}}, \\
{\cal W}_2 &=& -\frac{4i}{3} \frac{1}{{R_1R_2R_3}}\left[ R_1^2(2R_1^2 -R_2^2-R_3^2)
e^{12} - R_2^2(2R_2^2 - R_1^2-R_3^2)e^{34} + R_3^2(2R_3^2-R_1^2-R_2^2)e^{56} \right].\nn\\
\end{eqnarray}
$\bullet$ Data related to (\ref{data}): \bea m&=&1,
 \nn\\  n_1&=&n_2= 4,\nn\\
 \epsilon_1^{ab}&=&\delta^{ab}_{12},\nn\\
 \epsilon_2^{ab}&=&\delta^{ab}_{45},\nn\\\epsilon_3^{ab}&=&\delta^{ab}_{67},\nn\\
  \epsilon_1^{abc}&=&-\epsilon_2^{abc}=\epsilon_3^{abc}=\miso(\delta^{abc}_{146}+\delta^{abc}_{157}-\delta^{abc}_{247}+\delta^{abc}_{256}).  \eea
$\bullet$ Tables of field decompositions:

\

\begin{center}
\begin{tabular}{|c|c|c|}\hline \multicolumn{3}{|c|}{Table 5}\\
\multicolumn{3}{|c|}{Decomposition of bosonic fields under
$U(1)\times U(1)$}\\ \hline field & components & $U(1)\times
U(1)~\mbox{representations}$
\\
\hline \multirow{3}{*}{$\hat G_{MN}$} & $g_{\mu\nu}$ &
$\mathbf{(0,0)}$
\\
\cline{2-3} & $g_{\mu m}$ &
$\m{(0,2)}+\m{(0,-2)}+\m{(1,1)}+\m{(1,-1)}+\m{(-1,1)}+\m{(-1,-1)}$
\\
\cline{2-3} & \multirow{3}{*}{$g_{mn}$} &
$3\cdot\m{(0,0)}+\m{(0,4)}+\m{(0,-4)}+\m{(0,2)}+\m{(0,-2)}+ \m{(1,1)}+\m{(1,-1)}$ \\ \cline{3-3} & & $+\m{(1,3)}+\m{(1,-3)}+\m{(-1,1)}+\m{(-1,-1)}+\m{(-1,3)}+\m{(-1,-3)} $ \\ \cline{3-3} & & $+\m{(2,0)}+\m{(2,2)}+\m{(2,-2)}+\m{(-2,0)}+\m{(-2,2)}+\m{(-2,-2)}$ \\
\hline \multirow{3}{*}{$\hat B_{MN}$} & $B_{\mu\nu}$ & $\m{(0,0)}$
$$ \\ \cline{2-3} & $B_{\mu m}$ &
$\m{(0,2)}+\m{(0,-2)}+\m{(1,1)}+\m{(1,-1)}+\m{(-1,1)}+\m{(-1,-1)}$  \\ \cline{2-3}
& \multirow{2}{*}{$B_{mn}$} & $3\cdot\m{(0,0)}+\m{(0,2)}+\m{(0,-2)}+ \m{(1,1)}+\m{(1,-1)}+\m{(1,3)}+\m{(1,-3)}$ \\ \cline{3-3} & & $+\m{(-1,1)}+\m{(-1,-1)}+\m{(-1,3)}+\m{(-1,-3)}+\m{(2,0)}+\m{(-2,0)}$ \\
\hline
$\hat\phi$ & $\phi$ & $\mathbf{(0,0)}$ \\
\hline \multirow{2}{*}{$\hat A^{\hat A}_M$} & $A_\mu$ &
$\mathbf{(0,0)}$
\\ \cline{2-3}
& $A_m$ & $\m{(0,2)}+\m{(0,-2)}+\m{(1,1)}+\m{(1,-1)}+\m{(-1,1)}+\m{(-1,-1)}$ \\
\hline
\end{tabular}
\end{center}

\

\begin{center}
\begin{tabular}{|c|c|c|}\hline \multicolumn{3}{|c|}{Table 6}\\
\multicolumn{3}{|c|}{Decomposition of fermionic fields under
$U(1)\times U(1)$}\\ \hline field & components & $U(1)\times
U(1)~\mbox{representations}$ \\ \hline \multirow{2}{*}{$\hat\psi_M$} & $\psi_\mu$ &
$\m{(0,0)}+\m{(0,-1)}+\m{(1,1)}+\m{(-1,1)}$ \\
\cline{2-3}
& \multirow{2}{*}{$\psi_m$} & $3\cdot\m{(0,0)}+\m{(0,2)}+\m{(0,-2)}+\m{(0,4)}+3\cdot\m{(1,1)}+\m{(1,-1)}$ \\ \cline{3-3} & & $+\m{(1,3)}+\m{(1,-3)}+\m{(2,0)}+\m{(-2,0)}+\m{(2,-2)}+\m{(-2,-2)}$ \\
\hline
$\hat \lambda$ & $\lambda$ & $\m{(0,0)}+\m{(0,-1)}+\m{(1,1)}+\m{(-1,1)}$ \\
\hline
$\hat \chi$ & $\chi$ & $\m{(0,0)}+\m{(0,-1)}+\m{(1,1)}+\m{(-1,1)}$ \\
\hline
\end{tabular}
\end{center}

\end{document}